\documentclass[12pt,a4paper]{article}
\usepackage{amssymb,enumitem,booktabs,subfig,xcolor,todonotes,microtype,setspace}
\usepackage[top=1.4in, bottom=1.4in, left=1.35in, right=1.35in]{geometry}
\usepackage{natbib}
\usepackage{url} % not crucial - just used below for the URL
\usepackage[pdftex,colorlinks=true,hypertexnames=false]{hyperref}
\definecolor{darkblue}{rgb}{0,0,.6}
\hypersetup{citecolor=darkblue,linkcolor=darkblue,urlcolor=darkblue}
\usepackage{amsmath}
\usepackage{amsfonts}
\usepackage{epsfig}
\usepackage{graphics}
\usepackage{palatino,eulervm}
\usepackage{graphicx}
\usepackage{lscape}
\usepackage{rotating}
\usepackage[linewidth=1pt]{mdframed}
\allowdisplaybreaks[4]
\DeclareMathOperator*{\argmin}{arg\,min}

\providecommand{\U}[1]{\protect\rule{.1in}{.1in}}

\graphicspath{{plots/}}

\usepackage{amsthm,thmtools}
\usepackage{mathrsfs}
\declaretheorem{theorem}
\declaretheorem{lemma}
\makeatletter
\def\th@newremark{\th@remark\thm@headfont{\bfseries}}
\makeatletter
\theoremstyle{newremark}
\newtheorem{remark}{Remark}
\newtheorem{prop}{Proposition}

\newtheorem{assumption}{Assumption}
\declaretheoremstyle[
  spaceabove=6pt, spacebelow=6pt,
  headfont=\bfseries,
  notefont=\mdseries, notebraces={(}{)},
bodyfont=\normalfont,
  postheadspace=0.5em,
]{mystyle}

\begin{document}

\title{\LARGE Nonparametric Dyadic Quantile Regression}

\author{{\normalsize Francesco Bravo\thanks{Department of Economics and Related Studies, University of York, UK. Email: francesco.bravo@york.ac.uk.},\quad Degui Li\thanks{Faculty of Business Administration, Asia-Pacific Academy of Economics and Management, and Department of Economics, University of Macau, China. Email: deguili@um.edu.mo. Li's research is supported by the National Natural Science Foundation of China (72525015) and the CPG and SRG grants funded by the University of Macau.},\quad Zheng Li\thanks{Department of Agricultural and Resource Economics, North Carolina State University, US. Email: zli42@ncsu.edu. }}\\
{\normalsize\em University of York,\quad University of Macau,\quad North Carolina State University}}
\date{\normalsize This version: October 16, 2026}

\maketitle

\centerline{\bf Abstract}

\smallskip

This paper considers nonparametric quantile regression for directed dyadic data observed on the ordered pairs of a complete network with $n$ vertices. The observations exhibit dyadic dependence: dyads with disjoint indices are independent, whereas dyads sharing one or two indices are correlated. The local linear smoothing technique is adopted to estimate the quantile regression function. The asymptotic theory reveals that convergence rates of the proposed dyadic quantile regression estimation crucially depend on whether degeneracy occurs for the underlying network generating process. A rate-adaptive standardized version of the limit distribution is derived under some mild conditions. In addition, an easy-to-implement method is proposed to estimate the asymptotic variance of nonparametric dyadic quantile regression estimation. The finite-sample performance and applicability of the proposed methodology are illustrated with Monte-Carlo simulation and a real data application to bilateral portfolio holdings.

\medskip

\medskip


\newpage

%%%%%%%%%%%%%%%%%%%

\section{Introduction}\label{sec1}
\renewcommand{\theequation}{1.\arabic{equation}}
\setcounter{equation}{0}

The past two decades have seen notable interest and developments in network models with applications in various fields such as economics, finance, public health and psychology \citep[e.g.,][]{N02, DY15, S17, G20b, N21}. The primary interest of network data analysis often lies in exploring interconnections among a large number of units, with pairs of units forming dyads. Examples include international trade relationships between pairs of countries \citep{ST06} and \citep{AvW03}, informal risk sharing \citep{FG07}, strategic network formation \citep{G17}, friendships between pupils \citep{CFIK20}, policy choices between political units \citep{NP2010} and international financial flows \citep{PR2005} among many others. In these applications unobserved unit level characteristics, such as the multilateral resistance terms of \citet{AvW03} or specific households charcteristics, affect every dyad that involves a given unit, and ignoring this dependence typically produces standard errors that understate the sampling uncertainty \citep{CGM2011}. 

\smallskip

Let $i\in\{1,\cdots,n\}$ denote the index of a unit randomly drawn from a population, where $n$ is the number of units in the collected sample. Suppose we observe $X_i$, a vector associated with some characteristics of unit $i$, and $Y_{ij}$, a dyadic outcome associated with the edge between units $i$ and $j$. The aim of the present paper is to study the relationship between $Y_{ij}$ and $W_{ij}=(X_i^{^\intercal}, X_j^{^\intercal})^{^\intercal}$ for \textit{directed} dyadic data, that is for data in which the interaction between two dyadic members $i$ and $j$ initiates with $i$ and is directed towards $j$. For example, in international trade one can distinguish between the exporter and importer; and in international finance, one can distinguish between the home and foreign countries' investments. Unlike the classic regression analysis for independent cross-sectional data, both $Y_{ij}$ and $W_{ij}$ exhibit the so-called dyadic dependence, due to the paired nature of the data, i.e., $(Y_{ij}, W_{ij})$ and $(Y_{kl}, W_{kl})$ are correlated if they share a common unit index. This requires a different treatment and technique to derive the limit theory for dyadic regression models in contrast to the standard limit theory for classic regression models.      

\smallskip

Dyadic regression models have often been estimated under a pre-specified parametric structural assumption. \cite{CGM2011} and \cite{TM19} use ordinary least squares estimation and propose dyadic robust standard errors for inference; \cite{BCL11} propose a method of moments estimator; \cite{ST06} and \cite{G20a,G24} develop likelihood-based dyadic regression estimation and inference. Other recent developments and surveys on methodology, theory and algorithms for parametric dyadic regression can be found in \cite{DDG21}, \cite{GM21} and \cite{M21} and the references therein.

\smallskip

However, imposing a pre-specified structural assumption on the relationship between dyadic outcomes and some observed characteristics may be restrictive. When the model is misspecified \citep[e.g.,][]{G24},  dyadic regression estimation would result in biased estimators, implying that inference based on them would be invalid. This issue can be addressed by adopting a nonparametric approach, allowing the dyadic data to ``speak for themselves". Recently, there have been some attempts to nonparametrically estimate the dyadic density or regression function. \cite{GNP24} study kernel density estimation for undirected dyadic data and establish the point-wise consistency and asymptotic normality of the resulting kernel density estimator; \cite{CT23} derive uniform convergence rates for the kernel density estimator and propose a jackknife empirical likelihood approach for inference; \cite{CFU24} develop uniform inference (including minimax-optimal uniform convergence rates) for kernel density estimators; and \cite{GNP21} derive both point-wise and uniform convergence rates for the kernel-based dyadic nonparametric regression estimator.

\smallskip

It is well known that the conditional mean function cannot fully capture the impact of the observable characteristics $W_{ij}$ on the dyadic outcomes $Y_{ij}$, and nonparametric regression estimation may perform poorly when there are outliers. This paper addresses this important problem by developing a nonparametric dyadic quantile regression approach, which can be interpreted as a robust alternative to the nonparametric dyadic regression model \citep{GNP21}. Nonparametric quantile regression models for independent cross-sectional data and weakly dependent data have been considered by many authors, including  \cite{YJ98}, \cite{CX08}, \cite{CX12}, \cite{QY15}, \cite{BCCF19} and \cite{LLL21}, among many others. To the best of our knowledge, this is the first paper that considers a nonparametric quantile regression model for directed dyadic data with three main contributions summarized as follows. We first show that, because of the dyadic dependence, the relevant dimension for the developed local estimator is $d_X={\sf dim}(X_i)$ rather than $d_W={\sf dim}(W_{ij})=2d_X$ (for the ``non-degeneracy" scenario), which can be explained by the fact that the H\'ajek projection of local linear estimation in the context of dyadic regression has the so-called ``partial mean like" structure that reduces the dimension of conditioning variables, a phenomenon first explained by \cite{Newey1994} and noted by \cite{GNP21}. Second, we derive a rate-adaptive standardized asymptotic normal distribution theory, accounting for the scenario of ``degeneracy" that arises when the dependence across dyads vanishes. Third, we provide an easy-to-implement estimator for the asymptotic variance of the proposed nonparametric quantile regression estimator, which may be used to construct confidence intervals and more generally to conduct statistical inference. 

\smallskip

We conduct a Monte-Carlo study under both the non-degenerate and degenerate dyadic designs and find that the integrated mean squared errors contract at the two distinct rates revealed by the asymptotic theory, with the leave-two-node-out cross-validated bandwidth adapting to both regimes. The confidence intervals built from the proposed variance estimator attain coverage close to the nominal level in the degenerate design, while average coverage in the non-degenerate design moves toward the nominal level as the number of nodes increases. We then apply the developed methodology to bilateral portfolio holdings from the IMF Coordinated Portfolio Investment Survey, conditioned on country-level GDP, and find that it captures convex curvature and a heavy right tail without a logarithmic transformation, exposing a quantile-specific reporter-counterpart asymmetry that the conditional-mean regression would miss.

\smallskip

The rest of the paper is organized as follows. Section \ref{sec2} introduces the model and nonparametric estimation method for dyadic quantile regression. Section \ref{sec3} establishes the consistency and asymptotic normality of the proposed estimator together with assumptions and remarks, and discusses the asymptotic variance estimation methodology. Sections \ref{sec4} and \ref{sec5} report, respectively, the simulation study and empirical application. Section \ref{sec6} concludes the paper. The supplementary material contains the proofs of main results, technical lemmas with proofs as well as further discussion on practical issues in the variance estimation. %The following notation is used throughout the paper. We use $g^\prime$ to denote the derivative (or gradient) of function $g$, $^{^\intercal}$ to denote the transpose of a vector or matrix and $\sf{diag}\{a, b\}$ for a diagonal matrix with entries $a$ and $b$.

%%%%%%%%%%%%%%%%%%%

\section{Local linear dyadic quantile estimation}\label{sec2}
\renewcommand{\theequation}{2.\arabic{equation}}
\setcounter{equation}{0}

As in \cite{G20a} and \cite{GNP21}, we assume that $Y_{ij}$ is generated by a conditionally independent dyad (CID) model:
\begin{equation}\label{eq2.1}
Y_{ij}=h(X_i,X_j,U_i,U_j,V_{ij}),\quad 1 \leq i \neq j \leq n,
\end{equation} 
where $h$ is an unknown function which is allowed to be nonlinear and often called the graphon, $X_i$ and $U_i$ are independent and identically distributed (i.i.d.) random vectors over $i$, corresponding to the observed and unobserved features of unit $i$, and $V_{ij}$ are i.i.d. and independent of $\{X_i\}$ and $\{U_i\}$. The model setup (\ref{eq2.1}) implies the following properties of exchangeability \citep{Hoover1979, Aldous1981} and dyadic dependence: 

\begin{itemize}

\item Conditional on $W_{ij}=(X_i^{^\intercal},X_j^{^\intercal})^{^\intercal}$, $Y_{ij}$ are relatively exchangeable, i.e., for every permutation $\sigma$ satisfying $\{W_{\sigma(i)\sigma(j)}\}\stackrel{d}=\{W_{ij}\}$, we have $\{Y_{\sigma(i)\sigma(j)}\}\stackrel{d}=\{Y_{ij}\}$;

\item $Y_{ij}$ and $Y_{kl}$ are independent if their indices are disjoint, and may be correlated otherwise. 

\end{itemize}

\smallskip

Let $F_{Y|W}(\cdot|w)$ denote the conditional distribution function of the dyadic outcome $Y_{ij}$  given the vector of regressors $W_{ij}=w$. For $0<\tau<1$, define the conditional $\tau$-quantile function:
\begin{equation}\label{eq2.2}
{\sf Q}_{\tau}(w)=\inf\left\{y\in{\mathscr R}:\ F_{Y|W}(y|w)\geq \tau\right\}.
\end{equation} 
This is equivalent to 
\begin{equation}\label{eq2.3}
{\sf Q}_\tau(w)=\argmin_{y\in{\mathscr R}}{\sf E}\left[\rho_\tau(Y_{ij}-y)|W_{ij}=w\right],
\end{equation}
where $\rho_\tau(z)=z[\tau-I(z<0)]$ with $I(\cdot)$ being an indicator function. 

\smallskip

Assume that ${\sf Q}_\tau(\cdot)$ satisfies some smoothness condition, see Assumption \ref{ass:1} below, so that we can adopt the local linear smoothing technique of \cite{FG1996}, which is characterized by many good statistical properties over the local constant (Nadaraya-Watson) estimator, such as bias reduction, efficiency in the minimax sense and adaptivity near the boundary. We approximate ${\sf Q}_\tau(\cdot)$ by a Taylor expansion when $W_{ij}$ falls into a small neighborhood of $w$:
\[
{\sf Q}_\tau(W_{ij})\approx{\sf Q}_\tau(w)+(W_{ij}-w)^{^\intercal}{\sf Q}_\tau^\prime(w),
\] 
where ${\sf Q}_\tau^\prime(w)$ is a vector of the first-order derivatives of ${\sf Q}_\tau(w)$. 

\smallskip

Let $d_X$ be the dimension of $X_i$ and $d_W=2d_X$ be the dimension of $W_{ij}$. Define the weighted dyadic quantile regression loss function:
\begin{equation}\label{eq2.4}
L_{\tau}(a,b)=\frac{1}{n(n-1)}\sum_{1\leq i\neq j\leq n}\rho_\tau\left(Y_{ij}-a-(W_{ij}-w)^{^\intercal}b\right)K_h(W_{ij}-w),
\end{equation}
where $K_h(\cdot)=h^{-d_W}K(\cdot/h)$ with $K(\cdot)$ being a $d_W$-variate kernel function and $h$ being a bandwidth which will be determined by a leave-two-node-out cross validation, see Section \ref{sec4.2}. Minimizing $L_{\tau}(a,b)$ with respect to $a$ and $b$, we obtain the solution $\widehat{a}$ and $\widehat{b}$, and define the estimators of the quantile regression function and its derivative:
\begin{equation}\label{eq2.5}
\widehat{\sf Q}_\tau(w)=\widehat{a}\quad {\rm and}\quad\widehat{\sf Q}_\tau^\prime(w)=\widehat{b}.
\end{equation}
Note that apart from the summation taken over the index pairs $(i,j)$ of dyadic data, the loss function (\ref{eq2.4}) is virtually the same as that used, for example, by \cite{YJ98}, \cite{CX08} and \cite{LLL21} for independent or weakly dependent data. 

%%%%%%%%%%%%%%%%%%%%%%%%%%%%%%

\section{Asymptotic theory}\label{sec3}
\renewcommand{\theequation}{3.\arabic{equation}}
\setcounter{equation}{0}

In this section, we present the consistency and rate-adaptive standardized limit distribution theory for the developed nonparametric dyadic quantile regression estimation at a fixed quantile level $\tau$ and discuss the estimation of variance components.

\subsection{Consistency and asymptotic distribution theory}\label{sec3.1}

Let $f_{Y|W}\left(\cdot|\cdot\right)$ and $f_{W}\left(\cdot\right)$ denote, respectively, the conditional density of $Y_{ij}$ (given $W_{ij}$) and the density of $W_{ij}$. Write $w=(x_\circ^{^\intercal}, x_\bullet^{^\intercal})^{^\intercal}$ with both $x_\circ$ and $x_\bullet$ being $d_X$-dimensional. We start with some regularity conditions.

\smallskip

\begin{assumption}\label{ass:1}

{\em The conditional $\tau$-quantile function ${\sf Q}_{\tau}(w)$ defined in $(\ref{eq2.3})$ is unique and bounded with continuous derivatives up to the second order.}

\end{assumption}

\smallskip

\begin{assumption}\label{ass:2}

(i) {\em The marginal density function of $X_i$, $f_{X}(\cdot)$, is continuous and $f_{X}(x)>0$, $x=x_\circ, x_\bullet$.}

(ii) {\em The conditional density function of $Y_{ij}$ given $W_{ij}$, $f_{Y|W}(y | w')$, is bounded uniformly in $y\in\mathbb{R}$ and $w'$ in a neighborhood of $w$, is jointly continuous in $(y,w')$ on a neighborhood of $({\sf Q}_{\tau}(w),w)$, and satisfies $f_{Y|W}({\sf Q}_{\tau}(w)|w)>0$.}

%(iii) {\em there exist positive constants $C$ and $\epsilon$ such that $ f_{Y|W}\left(Q_{\tau}\left(w\right)+\eta\right) \leq C$ for all $|\eta| \leq \epsilon$.}

\end{assumption}

\smallskip

\begin{assumption}\label{ass:3}

(i) {\em The multivariate kernel function $K(\cdot)$ is continuous and symmetric with a compact support and in addition, $\int K(s)ds=1$.}

(ii) {\em Let the bandwidth $h$ satisfy $h\rightarrow0$ and $nh^{d_{X}} \rightarrow \infty$.}

\end{assumption}

\smallskip

\begin{remark}\label{re:1}

Assumptions \ref{ass:1}--\ref{ass:3} above are standard in the nonparametric local linear quantile regression literature \citep[e.g.,][]{FUT94, YJ98, QY15}. The uniqueness of ${\sf Q}_\tau(w)$ in Assumption 1 is  implied by Assumption 2(ii), since a positive conditional density in a neighborhood of ${\sf Q}_\tau(w)$ makes $F_{Y|W}(\cdot|w)$ strictly increasing there. The smoothness condition is satisfied by many specifications used in empirical work with dyadic data. For example, consider the multiplicative gravity model of \citet{ST06} that can be written in the CID form (\ref{eq2.1}):
\[
Y_{ij}=\exp\{g(X_i,X_j)\}\,\zeta_{ij},\qquad \zeta_{ij}=\exp(U_i+U_j)\,\eta_{ij},
\]
where $\{\eta_{ij}\}$ are positive, i.i.d.\ and independent of $\{X_i\}$ and $\{U_i\}$. Since $\zeta_{ij}$ is independent of $W_{ij}$, we have ${\sf Q}_\tau(w)=\exp\{g(w)\}\,q_\tau(\zeta)$ with $q_\tau(\zeta)$ being the $\tau$-quantile of $\zeta_{ij}$, so Assumption \ref{ass:1} holds whenever $g$ has continuous second-order derivatives. The shared node effects $U_i$ and $U_j$ make this a non-degenerate design. Due to the independence assumption on $\{X_i\}$, we have $f_W(w)=f_X(x_\circ)f_X(x_\bullet)$, hence Assumption \ref{ass:2}(i) indicates that $f_W(\cdot)$ is continuous and $f_W(w)>0$. Treating unit attributes as i.i.d. draws from a superpopulation is the standard sampling framework for dyadic data 
\citep{G20a,G20b}. When the units are countries, as in Section \ref{sec5}, this framework is best interpreted as describing sampling uncertainty about the network-generating process. The condition on  $K(\cdot)$ in Assumption \ref{ass:3}(i) facilitates our asymptotic proofs but the compact support restriction may be relaxed at the cost of slightly more lengthy proofs. The multivariate kernel can be constructed as 
\[K(s)=k(s_1)\cdots k(s_{d_W}),\quad s=(s_1,\cdots,s_{d_W})^{^\intercal},\] 
where $k(\cdot)$ is a univariate kernel function such as the Epanechnikov kernel. The bandwidth restriction in Assumption \ref{ass:3}(ii) is comparable to those in \cite{GNP21, GNP24}. \cite{GNP24} also consider the asymptotic distribution theory for kernel density estimation under the so-called ``knife-edge" smoothing condition \citep[e.g.,][]{CCJ14}.

\end{remark}

Define
\[
b_\tau(w)=\frac{1}{2}\int s^{^\intercal} \frac{\partial^{2}{\sf Q}_{\tau}(w)}{\partial w \partial w^{^\intercal}}s K(s)ds,\qquad
\sigma_n^2(w)=\frac{\nu_\tau(w)}{nh^{d_X}}+\frac{\nu_\tau^\dag(w)}{n(n-1)h^{d_W}},
\]
where 
\[
\nu_\tau(w)=\left(4h^{d_X}\right)\cdot\frac{Cov(T_{ik}^\circ(w),T_{jk}^\circ(w))}{(f_{Y|W}(\textsf{Q}_\tau(w))f_W(w))^2},\qquad \nu^\dag_\tau(w)=\frac{\tau(1-\tau)}{f_{Y|W}^2\left({\sf Q}_{\tau}(w)\right)f_{W}(w)}\int K^{2}(s) ds, 
\]
and 
\begin{equation}\label{eq3.1}
T_{ik}^\circ(w)=\frac{1}{2}\left[\psi_{\tau}(R_{ik})K_{h}(W_{ik}-w)+\psi_{\tau}(R_{ki})K_{h}(W_{ki}-w)\right]
\end{equation}
with $\psi_{\tau}(z)=I(z \leq 0) -\tau$ and $R_{ik}=Y_{ik}-{\sf Q}_{\tau}(W_{ik})$. The following theorem establishes the consistency and asymptotic normality of $\widehat{\sf Q}_\tau(w)$ defined in (\ref{eq2.5}).

\smallskip

\begin{theorem}\label{Th1}

Suppose that Assumptions \ref{ass:1}, \ref{ass:2} and \ref{ass:3}(i)(ii) are satisfied. 

(i) The local linear quantile regression estimate $\widehat{\sf Q}_\tau(w)$ exists and is weakly consistent. 

(ii) Suppose, in addition, $h^2=O(\sigma_n(w))$ and that either (a) $\nu_\tau(w)$ is bounded away from zero, or (b) the array $\{R_{ij}\}$ is i.i.d.\ and independent of $\{X_i\}$ and $\{U_i\})$ when $\nu_\tau(w)=0$. Then
 $\widehat{\sf Q}_\tau(w)$ has the following asymptotic distribution:
\begin{equation}\label{eq3.2}
\sigma_n^{-1}(w)\left[\widehat{\sf Q}_\tau(w)-{\sf Q}_\tau(w)-b_\tau(w)h^2\right]\stackrel{d}\longrightarrow {\sf N}\left(0, 1\right).
\end{equation}

\end{theorem}

\medskip

\begin{remark}\label{re:2}

(i) When $\nu_\tau(w)\neq0$, it follows from Theorem \ref{Th1}(ii) that the point-wise convergence rate of $\widehat{\sf Q}_\tau(w)$ is $h^2+(nh^{d_X})^{-1/2}$, where the first rate is due to the local linear estimation bias, the same as that in the existing literature \citep[e.g.,][]{CX08, LLL21}, whereas the second rate is due to the dominant term in the asymptotic variance $\sigma_n^2(w)$. In this non-degeneracy case, although the number of dyadic outcomes is $n(n-1)$, the effective sample size reflected in the convergence theory is $n$. This is due to the so-called ``dyadic-clustering" dependence \citep{M21} over dyadic outcomes sharing indices in common, which slows down the estimation convergence rate. Consequently, we have ``$n$" rather than ``$n(n-1)$" in the second rate. Meanwhile, the effective dimension for the kernel-based estimator is $d_X$. From (\ref{eq3.2}), the point-wise asymptotic mean squared error for $\widehat{\sf Q}_\tau(w)$ is
\[
{\sf E}\left\{\left[\widehat{\sf Q}_\tau(w)-{\sf Q}_\tau(w)\right]^2\right\}\sim b^2_\tau(w)h^4+\frac{\nu_\tau(w)}{nh^{d_X}}.
\]
Minimizing the right hand side with respect to $h$, we obtain an explicit expression for the optimal bandwidth:
\[
h_{\sf opt}= \left[\frac{d_X \nu_\tau(w)}{4b^2_\tau(w)}\right]^{1/(d_X+4)}n^{-1/(d_X+4)}=:c_{\sf opt}(w)n^{-1/(d_X+4)},
\]
which is comparable to that for kernel-weighted nonparametric estimation in the classic setup \citep[e.g.,][]{FY03}, treating $n$ as an effective sample size and $d_X$ as the relevant dimension (of covariates). Furthermore, with the optimal bandwidth, the point-wise mean squared error rate becomes $n^{-4/(d_X+4)}$.

(ii) The structure of $\sigma_n^{2}(w)$ allows for potential degeneracy of the underlying network generating process. Degeneracy arises when dependence across dyads vanishes. \cite{CFU24} and \cite{GNP24} consider kernel density estimation with dyadic data generated by
\[
W_{ij}^\dag=W(X_i,X_j,V_{ij}),
\]
and note that because ${\sf Var}(f_{W^\dag |X}(w^\dag|X_i))=0$, where $f_{W^\dag |X}(\cdot |\cdot)$ denotes the conditional density function of $W_{ij}^\dag$ given $X_i$ or $X_j$, the convergence rate and asymptotic distribution of the kernel estimator are different.  A similar phenomenon occurs in the context of nonparametric quantile regression estimation, which is illustrated via the following example for the case of an additive structure of the CID model (\ref{eq2.1}): 
\begin{equation}\label{eq3.3}
Y_{ij}=m(X_i,X_j)+V_{ij},
\end{equation}
where $m(\cdot,\cdot)$ is an unknown function, $\{X_i\}$ and $\{V_{ij}\}$ are defined as in Section \ref{sec2}, and, in addition, the $\tau$-quantile of $V_{ij}$ is zero. It is easy to verify that 
\[
m(x_\circ, x_\bullet)={\sf Q}_\tau(x_\circ, x_\bullet)={\sf Q}_\tau(w),\quad R_{ij}=Y_{ij}-{\sf Q}_{\tau}(W_{ij})=V_{ij},
\] 
and $\{R_{ij}\}$ is thus i.i.d. and independent of $\{X_i\}$. Then it is easy to see that $\nu_\tau(w)=0$, implying that the rate of the quantile estimation standard deviation is $[n(n-1)h^{d_W}]^{-1/2}$ whereas the asymptotic bias term remains the same, which is comparable to that for independent observations with sample size $n(n-1)$ and dimension of the covariates being $d_W$. Thus the standardized limit distribution in (\ref{eq3.2}) is rate adaptive, simultaneously accounting for two possible rates of convergence. 

(iii) Assumption \ref{ass:3}(ii) suffices for the estimation consistency, but the extra bandwidth restriction $h^2=O(\sigma_n(w))$ is required for the central limit theorem (\ref{eq3.2}) to tackle the linearization remainder of quantile regression estimation. Specifically, since ${\sf Q}_\tau(\cdot)$ is only assumed to be twice continuously differentiable, the smoothing bias beyond the leading term $b_\tau(w)h^2$ is $o(h^2)$, and the linearization remainder of the estimator is $o_P(h^2+\sigma_n(w))$, see Lemmas \ref{le:2} and \ref{le:5} in Appendix B. The extra bandwidth condition in Theorem \ref{Th1}(ii) makes both of these $o_P(\sigma_n(w))$. This condition holds at the optimal bandwidth order $h\asymp n^{-1/(d_X+4)}$ under non-degeneracy and $h\asymp n^{-1/(d_X+2)}$ under degeneracy, for which $h^2\asymp\sigma_n(w)$ in each case, and it holds under any undersmoothing such as the rule used in Section \ref{sec4.4}, for which $h^2=o(\sigma_n(w))$.

%{\color{red} In fact, the distribution result (\ref{eq3.2}) may fail without this bandwidth condition. For example, let $d_X=1$ and the degenerate model (\ref{eq3.3}) hold with $\tau=1/2$ and $V_{ij}\sim U(-1,1)$, and suppose that $m(\cdot)$ is quartic near $w=(0,0)^{^\intercal}$ with a covariate density that is flat near zero. Then $b_\tau(w)=0$ and the smoothing bias is of exact order $h^4$, whereas $\sigma_n(w)\asymp(nh)^{-1}$. A bandwidth $h=n^{-1/10}$ satisfies Assumption \ref{ass:3}(ii), yet the bias-to-standard-error ratio diverges at rate $n^{1/2}$. Assumption \ref{ass:3}(iii) is mild. }

\end{remark}

\subsection{Estimation of the variance components}\label{sec3.2}

We next propose an estimator for the asymptotic variance $\sigma_n^2(w)$, which is needed for the construction of point-wise confidence intervals and statistical inference. Since the asymptotic variance contains two different components, we consider their construction separately.
 
\subsubsection{Estimation of $\nu_\tau^\dag(w)$}\label{sec3.2.1}

It follows from the definition of $\nu_\tau^\dag(w)$ that we only need to consistently estimate the density function $f_W(w)$ and the conditional density function $f_{Y|W}({\sf Q}_\tau(w))$. With the standard kernel estimation method, we can consistently estimate $f_W(w)$ as in \cite{CFU24} and \cite{GNP24}, and denote the resulting estimate as $\widehat{f}_W(w)$.

\smallskip

We next turn to the estimation of $f_{Y|W}({\sf Q}_\tau(w))$ using the kernel regression method introduced by \cite{FYT96}. Assuming that ${\sf Q}_\tau(w)$ is known, by some standard calculation, we obtain 
\begin{equation}\label{eq3.4}
f_{Y|W}({\sf Q}_\tau(w))\approx {\sf E}\left[k_{b_1}(Y_{ij}-{\sf Q}_\tau(w))|W_{ij}=w\right]\quad \text{as}\quad b_1\rightarrow0,
\end{equation}
where $k_{b_1}(\cdot)=b_1^{-1}k(\cdot/b_1)$ with $k(\cdot)$ being a univariate kernel function. By (\ref{eq3.4}), we may consistently estimate $f_{Y|W}({\sf Q}_\tau(w))$ by the conventional nonparametric regression method, treating $k_{b_1}(Y_{ij}-{\sf Q}_\tau(w))$ as the response and $W_{ij}=(X_i^{^\intercal}, X_j^{^\intercal})^{^\intercal}$ as the covariate. Specifically, we consider the following local objective function:
\[
\frac{1}{n(n-1)}\sum_{1\leq i\neq j\leq n}\left[k_{b_1}(Y_{ij}-{\sf Q}_\tau(w))-c\right]^2H_{b_2}(W_{ij}-w),
\] 
where $H_{b_2}(\cdot)=b_2^{-d_W}H(\cdot/b_2)$ with $H(\cdot)$ being a multivariate kernel function and $b_2$ being a bandwidth. Minimizing the above objective function with respect to $c$, we obtain
\begin{equation}\label{eq3.5}
\widetilde{f}_{Y|W}({\sf Q}_\tau(w))=\frac{\sum_{1\leq i\neq j\leq n}k_{b_1}(Y_{ij}-{\sf Q}_\tau(w))H_{b_2}(W_{ij}-w)}{\sum_{1\leq i\neq j\leq n}H_{b_2}(W_{ij}-w)}.
\end{equation}
However, the conditional density function estimation in (\ref{eq3.5}) is practically infeasible as ${\sf Q}_\tau(w)$ is unknown. To address this issue and facilitate the development of asymptotic theory, we next adopt a sample-splitting technique introduced by \cite{LLL21} for bandwidth selection in the nonparametric quantile regression estimation. We split the full sample into two blocks:
\[
{\mathscr M}_1=\{(Y_{ij}, W_{ij}): 1\leq i\neq j\leq n_1\},\quad {\mathscr M}_2=\{(Y_{ij}, W_{ij}): n_1+1\leq i\neq j\leq n\},
\]
where $n_1\propto n$. For simplicity, we may set $n_1=\lfloor n/2\rfloor$. With the observations in ${\mathscr M}_1$, we run the dyadic quantile regression as in Section \ref{sec2} and obtain the resulting estimator $\widehat{\sf Q}_{\tau}^{{\mathscr M}_1}(w)$. Due to the independence assumption in $\{X_i\}$, $\{U_i\}$ and $\{V_{ij}\}$, $\widehat{\sf Q}_{\tau}^{{\mathscr M}_1}(w)$ is independent of ${\mathscr M}_2$. We then construct a feasible conditional density function estimation: 
\begin{equation}\label{eq3.6}
\widehat{f}_{Y|W}({\sf Q}_\tau(w))=\frac{\sum_{n_1+1\leq i\neq j\leq n}k_{b_1}(Y_{ij}-\widehat{\sf Q}_\tau^{{\mathscr M}_1}(w))H_{b_2}(W_{ij}-w)}{\sum_{n_1+1\leq i\neq j\leq n}H_{b_2}(W_{ij}-w)}.
\end{equation}
We require the following conditions on kernel functions and bandwidths to derive the consistency property of $\widehat{f}_{Y|W}({\sf Q}_\tau(w))$.

%\begin{assumption}\label{ass:4}

% {\em The conditional density function $f_{Y|W}(\cdot|w)$ is twice continuously differentiable with respect to ${\sf Q}_\tau(w)$  with second derivative bounded at a neighborhood of ${\sf Q}_\tau(w')$ for $ w'$ in a neighborhood of $w$.}

%\end{assumption}

\begin{assumption}\label{ass:4}

(i) {\em The univariate kernel function $k(\cdot)$ is symmetric with bounded first-order derivative $\dot k(\cdot)$ satisfying $\int |\dot k(u)|\,du < \infty$}.

(ii) {\em The multivariate kernel function $H(\cdot)$ is symmetric, bounded with compact support.}

\end{assumption}

\begin{assumption}\label{ass:5}

\em{The bandwidths $b_1$ and $b_2$ satisfy $b_1, b_2 \rightarrow 0$, 
\[
nb_1\to\infty,\quad nb_2^{d_W} \rightarrow \infty\quad \text{and}\quad h^2+(nh^{d_X})^{-1/2}=o(b_1).
\]
}

\end{assumption}

\smallskip

Proposition \ref{prop:1} below gives the consistency property of conditional density function estimation via sample splitting in the context of dyadic regression, which is of independent interest.

\smallskip

\begin{prop}\label{prop:1}

{\em Suppose that Assumptions $\ref{ass:1}$--$\ref{ass:5}$ are satisfied. The following weak consistency property holds,
\begin{equation}\label{eq3.7}
\left\vert \widehat{f}_{Y|W}({\sf Q}_\tau(w))-f_{Y|W}({\sf Q}_\tau(w))\right\vert=o_P(1).
\end{equation}
}

\end{prop}

\smallskip

With Proposition \ref{prop:1}, a consistent estimator of $\nu_\tau^\dag(w)$ is defined as
\begin{equation}\label{eq4.8}
\widehat\nu_\tau^\dag(w)=\frac{\tau(1-\tau)}{\widehat{f}_{Y|W}^2\left({\sf Q}_{\tau}(w)\right)\widehat{f}_{W}(w)}\int K^{2}(s) ds.
\end{equation}

\subsubsection{Estimation of $\nu_\tau(w)$}\label{sec3.2.2}

We next consider estimating $\nu_\tau(w)$ for the non-degeneracy case. Since $f_{Y|W}({\sf Q}_\tau(w))$ and $f_W(w)$ can be consistently estimated as in Section \ref{sec3.2.1}, it is sufficient to estimate ${\sf Cov}\left(T_{ik}^\circ(w),T_{jk}^\circ(w)\right)$, $i\neq j\neq k$. It follows from the proof of Theorem \ref{Th1} that 
\begin{equation}\label{eq3.9}
{\sf Cov}\left(T_{ik}^\circ(w),T_{jk}^\circ(w)\right)={\sf Var}\left( A_k^\circ(w)\right),\quad A_k^\circ(w)={\sf E}\left(T^\circ_{k1}(w)|X_k, U_k\right),
\end{equation}
and 
\begin{equation}\label{eq3.10}
\frac{\frac{1}{n}\sum_{k=1}^n\left[ A_k^\circ(w) \right]^2-{\sf Var}\left( A_1^\circ(w)\right)}{{\sf Var}\left( A_1^\circ(w)\right)}=o_P(1),
\end{equation}
see (\ref{eqA.8}) and (\ref{eqA.13}) in Appendix A. By virtue of (\ref{eq3.9}) and (\ref{eq3.10}), we obtain a natural estimator of ${\sf Cov}\left(T_{ik}^\circ(w),T_{jk}^\circ(w)\right)$:
\begin{equation}\label{eq3.11}
\widetilde{\sf Cov}\left(T_{ik}^\circ(w),T_{jk}^\circ(w)\right)=\frac{1}{n}\sum_{k=1}^n\left[ A_k^\circ(w) \right]^2.
\end{equation}

\smallskip

However, the above estimator is infeasible as $A_k^\circ(w)$ is unobservable. We next adopt the sample-splitting technique again as in Section \ref{sec3.2.1} to obtain the approximation to $A_k^\circ(w)$. Define
\[
\widehat{R}_{ij}=Y_{ij}-\widehat{\sf Q}_\tau^{{\mathscr M}_1}(w),\quad \widehat{T}_{ij}^\circ(w)=\frac{1}{2}\left[ \psi_\tau(\widehat{R}_{ij})K_h(W_{ij}-w)+\psi_\tau(\widehat{R}_{ji})K_h(W_{ji}-w)\right].
\]
Consequently, we approximate $A_k^\circ(w)$ by
\[
\widehat{A}_k^\circ(w)=\frac{1}{n-n_1-1}\sum_{n_1+1\leq j\leq n, j\neq k}\widehat{T}_{jk}^\circ(w)
\]
and estimate ${\sf Cov}\left(T_{ik}^\circ(w),T_{jk}^\circ(w)\right)$ for pairwise distinct $i,j$ and $k$, by
\begin{equation}\label{eq3.12}
\widehat{\sf Cov}\left(T_{ik}^\circ(w),T_{jk}^\circ(w)\right)=\frac{1}{n-n_1}\sum_{k=n_1+1}^n\left[ \widehat{A}_k^\circ(w) \right]^2.
\end{equation}
Finally, we estimate $\nu_\tau(w)$ by
\begin{equation}\label{eq3.13}
\widehat\nu_\tau(w)=4h^{d_X}\cdot\frac{\widehat{\sf Cov}\left(T_{ik}^\circ(w),T_{jk}^\circ(w)\right)}{\left[\widehat f_{Y|W}({\sf Q}_\tau(w))\widehat f_W(w)\right]^2}.
\end{equation}

%%%%%%%%%%%%%%%%%%%

\section{Monte-Carlo simulation}\label{sec4}
\renewcommand{\theequation}{4.\arabic{equation}}
\setcounter{equation}{0}

This section evaluates the finite-sample performance of the proposed methodology under Gaussian and standardized Student $t(5)$ disturbances. It is divided into four subsections: Section~\ref{sec4.1} introduces the simulation setup and data generating process (DGP); Section \ref{sec4.2} discusses a data-driven bandwidth selection criterion; Section \ref{sec4.3} examines the estimation accuracy of the local linear dyadic quantile estimation, illustrating the rate-adaptive limiting theory in Section \ref{sec3.1}; and Section~\ref{sec4.4} assesses the finite-sample performance of the asymptotic variance estimation discussed in Section~\ref{sec3.2}, as measured by the coverage of point-wise confidence intervals it delivers. 

\subsection{Simulation design}\label{sec4.1}

Across all experiments, we generate $\{X_i\}_{i=1}^n$ independently with $X_i\sim\mathrm{Unif}(0,1)$. The nonlinear regression function is specified as
\begin{equation}\label{eq4.1}
m(x_\circ,x_\bullet)=\log(1+2x_\circ+x_\bullet)+0.5\,x_\circ x_\bullet,
\end{equation}
which is asymmetric, since in general $m(x_\circ,x_\bullet)\neq m(x_\bullet,x_\circ)$, allowing the design to reflect the directed nature of dyadic interaction.

\smallskip

We consider two Gaussian DGPs that share the same regression function (\ref{eq4.1}) and differ only through the dependence structure of the disturbances. In the non-degeneracy design, define
\[
Y_{ij}=m(X_i,X_j)+U_i+U_j+V_{ij},\qquad 1\leq i\neq j\leq n,
\]
where
\[
U_i\stackrel{\mathrm{i.i.d.}}{\sim}{\sf N}(0,1/4),
\qquad
V_{ij}\stackrel{\mathrm{i.i.d.}}{\sim}{\sf N}(0,1/2),
\]
and $\{X_i\}$, $\{U_i\}$, and $\{V_{ij}\}$ are mutually independent. Conditional on $W_{ij}=w=(x_\circ, x_\bullet)^{^\intercal}$, the composite disturbance $U_i+U_j+V_{ij}$ is Gaussian with variance one. Hence the conditional $\tau$-quantile function is given by ${\sf Q}_\tau(w)=m(x_\circ,x_\bullet)+\Phi^{-1}(\tau)$, where $\Phi^{-1}(\cdot)$ denotes the standard normal quantile function. Because dyads that share a unit also share one of the latent features, the first-order projection is nonzero, equivalently, $\nu_\tau(w)\neq 0$ in the notation of Section~\ref{sec3}, representing the case of non-degeneracy in the asymptotic theory.

\smallskip

In the design with degeneracy, we construct $Y_{ij}$ using the model (\ref{eq3.3}), where $V_{ij}\stackrel{\mathrm{i.i.d.}}{\sim}{\sf N}(0,1)$ is independent of $\{X_i\}$. The conditional $\tau$-quantile function is again ${\sf Q}_\tau(w)=m(x_\circ,x_\bullet)+\Phi^{-1}(\tau)$. Because the disturbance depends only on the dyad-specific shock $V_{ij}$, the first-order projection vanishes, $\nu_\tau(w)=0$, corresponding to the degenerate case.

\smallskip

To assess the sensitivity of these results to the error distribution, we also consider
standardized Student $t(5)$ disturbances under each dependence structure. Let $G_{ij}$ denote
the Gaussian disturbance in the corresponding design above, namely $U_i+U_j+V_{ij}$ in the
non-degenerate design and $V_{ij}$ in the degenerate design. We generate
\[
Y_{ij}=m(X_i,X_j)+\sqrt{\frac35}\,F_5^{-1}\!\left(\Phi(G_{ij})\right),
\qquad 1\leq i\neq j\leq n,
\]
where $F_5$ is the Student $t(5)$ distribution function and $\Phi$ is the standard normal
distribution function. The transformation retains dependence through shared nodes in the
non-degenerate design and independence of the disturbances across dyads in the degenerate
design. The factor $\sqrt{3/5}$ standardizes the error variance to one, matching the Gaussian
experiments. Consequently, differences between the Gaussian and Student results reflect
distributional shape rather than error scale. The conditional quantile function in the
Student designs is
\[
{\sf Q}_\tau(w)=m(x_\circ,x_\bullet)+\sqrt{\frac35}\,F_5^{-1}(\tau).
\]
The regression function, covariate distribution and network structure are otherwise unchanged,
allowing a direct comparison of estimation accuracy across the two error distributions.

\smallskip

In the Monte-Carlo experiment, the number of nodes $n$ takes values in $\{15,30,60\}$. These complete directed networks contain $n(n-1)$ observed dyads, corresponding to 210, 870 and 3,540 data points, respectively. The quantile index $\tau$ takes values in $\{0.10,0.25,0.50,0.75,0.90\}$. In all cases, the number of Monte-Carlo replications is $r=1000$. The local linear dyadic quantile estimator is implemented with the product Gaussian kernel
\[
K_h(W_{kl}-w)=\frac{1}{h^2}\,
\phi\!\left(\frac{X_k-x_\circ}{h}\right)
\phi\!\left(\frac{X_l-x_\bullet}{h}\right),
\qquad
\phi(z)=\frac{1}{\sqrt{2\pi}}\,e^{-z^2/2}.
\]
For each design and each quantile level, we summarize the finite-sample estimation accuracy by computing the integrated mean squared error, which for complete dyad-level fits is
\[
\mathrm{IMSE}_{\tau}=\frac{1}{r}\sum_{\ell=1}^{r}\frac{1}{n(n-1)}\sum_{1\leq i\neq j\leq n}\Bigl[\widehat{\sf Q}_{\tau,\ell}(W_{ij,\ell})-{\sf Q}_\tau(W_{ij,\ell})\Bigr]^2,
\]
where $W_{ij,\ell}=(X_{i,\ell},X_{j,\ell})^{^\intercal}$ denotes the covariate vector for dyad $(i,j)$ in the $\ell$-th replication. The squared errors are averaged over the dyadic covariate values in each replication and then across replications. Within each error distribution, the two dependence designs share the same population quantile regression function, so differences in IMSE reflect their distinct dyadic dependence structures. Comparing the Gaussian and standardized Student designs then shows how estimation accuracy varies with the shape of the disturbance distribution.

\subsection{Leave-two-node-out cross validation}\label{sec4.2}

The asymptotic theory in Section~\ref{sec3} shows that the bandwidth rate that minimizes the asymptotic mean squared error depends on whether the variance term $\nu_\tau(w)$ vanishes: the optimal bandwidth shrinks at the rate $n^{-1/(d_X+4)}$ in the non-degeneracy case, i.e., $\nu_\tau(w)\neq 0$,
and at the faster rate $n^{-1/(d_X+2)}$ when degeneracy occurs, i.e., $\nu_\tau(w)=0$. In practical applications, there is no prior information on which regime to apply since $\nu_\tau(w)$ is an unknown population quantity. A data-driven bandwidth selection criterion that responds to the dependence in the sample, without requiring the analyst to commit to one of the two rates ex ante, is therefore desirable.

\smallskip

In each Monte-Carlo replication, we determine the bandwidth by the leave-two-node-out (LTNO) cross validation. Let ${\mathscr D}_n=\{\{i,j\}:1\leq i<j\leq n\}$ denote the collection of unordered pairs of nodes, with $|\mathscr D_n|=\binom{n}{2}$. For each pair $\{i,j\}\in\mathscr D_n$, the validation set contains the two ordered dyads $(i,j)$ and $(j,i)$, whereas the training set is
\[
\mathscr T_{\{i,j\}}=\bigl\{(k,l):\,1\leq k\neq l\leq n,\,k\notin\{i,j\}\text{ and }l\notin\{i,j\}\bigr\}.
\]
Let $\widehat{\sf Q}_{\tau,-\{i,j\},h}(\cdot)$ denote the local linear dyadic quantile estimate fitted on $\mathscr T_{\{i,j\}}$ with bandwidth $h$. The LTNO criterion is
\[
{\sf CV}_\tau(h)=\frac{1}{n(n-1)}\sum_{\{i,j\}\in\mathscr D_n}\left[\rho_\tau\!\left(Y_{ij}-\widehat{\sf Q}_{\tau,-\{i,j\},h}(W_{ij})\right)+\rho_\tau\!\left(Y_{ji}-\widehat{\sf Q}_{\tau,-\{i,j\},h}(W_{ji})\right)\right],
\]
which averages the check loss over the $n(n-1)$ ordered dyads. The bandwidth is determined by
\[
\widehat h_{\rm cv}=\argmin_{h>0}{\sf CV}_\tau(h).
\]

\smallskip

The standard leave-one-dyad-out (LOO) rule, which drops a single ordered dyad $(i,j)$ at a time, is not suitable for dyadic data with a non-zero first-order projection. This is due to the fact that the held-out dyad $(i,j)$ shares its two endpoints with many other dyads $(i,k)$ and $(k,j)$ in the training set, so the node-level (latent) random effects $U_i$ and $U_j$ are still identifiable from the training set even though the validation point itself is absent. Under the Gaussian non-degenerate DGP, where the
disturbance contains an additive $U_i+U_j$ component, the kernel-weighted local fit at $w=W_{ij}$ can therefore implicitly recover $U_i$ and $U_j$ from those neighboring training dyads. The smaller the bandwidth, the more concentrated the kernel weight on dyads that share an endpoint with $(i,j)$, and the sharper this recovery becomes. As a result, the LOO criterion mistakenly rewards bandwidths that exploit information about the held-out node effects, and the data-driven
$\widehat h_{cv}$ is biased downward in the non-degenerate design. Indeed, we observe that, under the non-degenerate DGP, the LOO rule tends to select smaller bandwidths than the degenerate DGP, contrary to the theoretical ordering of optimal rates. In contrast, the proposed LTNO removes both endpoints of the validation dyad from the training set. Because no training dyad involves either node $i$ or node $j$, neither $U_i$ nor $U_j$ leaves any footprint in $\mathscr T_{\{i,j\}}$, and the kernel-weighted training fit cannot recover them at the validation point. The check loss on the held-out pair then reflects the honest extrapolation error against the
unknown components of the conditional distribution, in both the non-degenerate and degenerate DGPs. The argument above is intuitive only, and a rigorous theoretical analysis for the optimality of the LTNO bandwidth selection in the dyadic quantile regression will be left to future work. We use the same LTNO criterion in both dependence designs and under both error distributions, so that all comparisons use the same bandwidth selection principle.

%Applying the same data-driven rule in both designs ensures that the comparison between the non-degenerate and degenerate cases is conducted on equal footing, and that the resulting bandwidths can be interpreted as data-driven counterparts of the optimal asymptotic rates of Section~\ref{sec3}.

\subsection{Quantile regression estimation accuracy}\label{sec4.3}

Using the theoretically optimal bandwidth orders discussed in Section \ref{sec4.2}, the asymptotic theory in Section~\ref{sec3.1} gives $\mathrm{IMSE}\propto n^{-4/(d_X+4)}=n^{-4/5}$ in the non-degenerate case and $\mathrm{IMSE}\propto n^{-4/(d_X+2)}=n^{-4/3}$ in the degenerate case noting that $d_X=1$ in the simulation. Doubling the number of units $n$ should therefore divide IMSE by $2^{4/5}\approx 1.74$ in the non-degenerate design and by $2^{4/3}\approx 2.52$ in the degenerate design.

\smallskip

Table~\ref{tab:dyadic-imse-all} reports the Monte-Carlo IMSE (multiplied by $1000$) for all five quantile levels under both dependence designs, both error distributions and each of the three sample sizes. Consider first the Gaussian results. In both designs and for every value of $\tau$, IMSE decreases as $n$ increases, and the decreasing rate is systematically faster in the degenerate design than that in the non-degenerate design. At $\tau=0.50$, for example, doubling $n$ from $30$ to $60$ divides IMSE by roughly $1.9$ in the non-degenerate design and by roughly $3.2$ in the degenerate design, compared with the theoretical contractions $1.74$ and $2.52$. At the smaller level of doubling from $n=15$ to $n=30$, the observed contractions are about $1.9$ and $3.4$, respectively. The finite-sample reductions are thus larger than the asymptotic benchmarks, particularly in the degenerate design. Overall, the observed contractions are broadly consistent with the difference in theoretical convergence speeds. At the median, the contractions from $n=30$ to $n=60$ are closer to the asymptotic benchmarks than those from $n=15$ to $n=30$.

\smallskip

For the Gaussian experiments, the cross-design comparison is also sharp: at every sample size and quantile level, the non-degenerate IMSE is several times larger than degenerate IMSE, by a factor of about $3$ to $4$ at $n=15$ that grows to about $8$ to $13$ at $n=60$. This ratio increases with $n$, reflecting the distinct convergence speeds between the two designs over the sample sizes considered.

\smallskip

The standardized Student experiments exhibit the same ordering across dependence designs.
IMSE decreases with $n$ at every quantile and remains larger in the non-degenerate design.
At $\tau=0.50$, increasing $n$ from 30 to 60 reduces IMSE from 48.14 to 26.26 in the
non-degenerate design and from 7.04 to 2.09 in the degenerate design, with all values scaled
by 1000. The corresponding contraction factors, approximately 1.8 and 3.4, are similar to
their Gaussian counterparts. At $n=60$, non-degenerate IMSE is roughly 8 to 13 times its
degenerate counterpart across the five quantiles. The effect of dyadic dependence on estimation
accuracy is therefore evident under both error distributions.

\smallskip

Comparing the two error distributions reveals a different pattern across quantiles. At each
of the three sample sizes, standardized Student errors produce lower IMSE at the three central
quantiles $\tau=0.25,0.50,0.75$, but higher IMSE at $\tau=0.10,0.90$, in both dependence designs.
For example, at $n=60$ and $\tau=0.50$, IMSE falls from 39.35 to 26.26 in the non-degenerate
design and from 3.09 to 2.09 in the degenerate design when Gaussian errors are replaced by
standardized Student errors. At $\tau=0.10$, the corresponding changes are from 45.38 to
52.66 and from 5.27 to 5.86. With the error variance held fixed, heavier tails thus increase
estimation error at the outer quantiles without producing a uniform deterioration in accuracy.

\begin{table}[!htbp]
\centering
\caption{Monte-Carlo IMSE ($\times 1000$): Gaussian and standardized $t(5)$ designs}
\label{tab:dyadic-imse-all}
\begin{tabular}{lrrrrrr}
\toprule
& \multicolumn{3}{c}{Gaussian} & \multicolumn{3}{c}{Standardized $t(5)$} \\
\cmidrule(lr){2-4}\cmidrule(lr){5-7}
$\tau$ & $n=15$ & $n=30$ & $n=60$ & $n=15$ & $n=30$ & $n=60$ \\
\midrule
\multicolumn{7}{l}{\textit{Panel A: Non-degenerate}} \\
$0.10$ & 177.76 & 88.57 & 45.38 & 193.93 & 98.34 & 52.66 \\
$0.25$ & 144.59 & 73.17 & 39.96 & 123.99 & 58.38 & 29.56 \\
$0.50$ & 139.03 & 73.30 & 39.35 & 99.18 & 48.14 & 26.26 \\
$0.75$ & 149.02 & 77.73 & 41.37 & 120.51 & 59.51 & 32.72 \\
$0.90$ & 189.23 & 90.73 & 43.51 & 192.30 & 98.86 & 51.61 \\
\midrule
\multicolumn{7}{l}{\textit{Panel B: Degenerate}} \\
$0.10$ & 62.11 & 18.22 & 5.27 & 72.08 & 21.34 & 5.86 \\
$0.25$ & 39.68 & 12.19 & 3.50 & 32.20 & 9.23 & 2.73 \\
$0.50$ & 33.75 & 9.95 & 3.09 & 23.38 & 7.04 & 2.09 \\
$0.75$ & 41.31 & 11.58 & 3.54 & 30.93 & 9.08 & 2.74 \\
$0.90$ & 63.28 & 18.27 & 5.44 & 72.35 & 21.20 & 6.15 \\
\bottomrule
\end{tabular}
\par\smallskip
\begin{minipage}{0.94\textwidth}
\footnotesize Notes: Squared errors are averaged over dyad-level fits within each replication and then
across replications. Bandwidths are selected by LTNO cross validation. Both error distributions have variance one. Smaller values
indicate better finite-sample performance.
\end{minipage}
\end{table}

\subsection{Performance of the variance estimation}\label{sec4.4}

We next turn to the finite-sample performance of the asymptotic variance estimator of Section~\ref{sec3.2}, as measured by the coverage of the point-wise confidence intervals. This exercise uses the two Gaussian DGPs in Section~\ref{sec4.1}, together with the product Gaussian kernel and the LTNO bandwidth selection criterion in Section~\ref{sec4.2}. The evaluation point is fixed at $w=(0.5,0.5)^{^\intercal}$, an interior point of the support of $W_{ij}$. For each design, each $n\in\{15,30,60\}$, and each $\tau\in\{0.10,0.25,0.50,0.75,0.90\}$, in each replication we compute $\widehat{\sf Q}_\tau(w)$, assemble the plug-in variance estimate $\widehat\sigma_n^2(w)$, and record whether the nominal $95\%$ interval $\widehat{\sf Q}_\tau(w)\pm 1.96\,\widehat\sigma_n(w)$ covers the true quantile ${\sf Q}_\tau(w)$. For interval construction, the smoothing bias must be small relative to the standard error. A bandwidth chosen for point estimation balances the leading squared bias and variance, so using it without adjustment can leave a non-negligible bias in the normal approximation. We therefore undersmooth the regression estimate by setting
\[
\widehat h_{\rm US}=\widehat h_{\rm cv}\,n^{-1/20},
\]
following the multiplicative rule used by \cite{LLL21}. When $\widehat h_{\rm cv}$ has the relevant MSE-optimal order, the additional shrinkage makes the leading local linear bias, which is of order $h^2$, asymptotically negligible relative to the standard error under either dependence regime. In particular, $\widehat h_{\rm US}$ then satisfies $h^2=o(\sigma_n(w))$ and the estimation bias term may be omitted from the interval by Theorem \ref{Th1}(ii). In finite samples, the rule reduces smoothing bias without requiring an explicit estimate of the bias function. The quantile and variance estimates are computed at $\widehat h_{\rm US}$.

\smallskip

The auxiliary density bandwidths $b_Y$ and $b_W$, corresponding to $b_1$ and $b_2$ in (\ref{eq3.6}), are selected jointly by LTNO likelihood cross validation. Specifically, we minimize the negative conditional log likelihood
\[
{\sf CV}_f(b_Y,b_W)
=-\frac{1}{n(n-1)}\sum_{i\neq j}
\log\widehat f_{Y|W,-\{i,j\};b_Y,b_W}(Y_{ij}\mid W_{ij}),
\]
where $\widehat f_{Y|W,-\{i,j\};b_Y,b_W}$ is estimated after excluding all dyads involving either node $i$ or node $j$. This criterion evaluates the predictive density at the observed outcome of each validation dyad while respecting the dyadic dependence structure. The density bandwidths are retained when the regression bandwidth is undersmoothed, so quantile estimation and density estimation are tuned to their respective objectives.

\smallskip

Appendix C in the supplement describes the diagnostic used to choose between the variance components in $\widehat\sigma_n^2(w)$. In this experiment, we apply the diagnostic around $\widehat h_{\rm cv}$.

\smallskip

\begin{table}[!htbp]
\centering
\caption{Empirical coverage of nominal $95\%$ confidence intervals}
\label{tab:dyadic-cover-all}
\begin{tabular}{lrrr}
\toprule
$\tau$ & $n=15$ & $n=30$ & $n=60$ \\
\midrule
\multicolumn{4}{l}{\textit{Panel A: Non-degenerate}} \\
$0.10$ & 85.90 & 87.40 & 88.90 \\
$0.25$ & 90.00 & 91.20 & 91.20 \\
$0.50$ & 91.90 & 92.10 & 92.10 \\
$0.75$ & 87.60 & 88.00 & 91.30 \\
$0.90$ & 83.40 & 87.50 & 87.70 \\
\midrule
\multicolumn{4}{l}{\textit{Panel B: Degenerate}} \\
$0.10$ & 93.40 & 96.10 & 96.70 \\
$0.25$ & 97.10 & 97.90 & 96.40 \\
$0.50$ & 97.10 & 97.30 & 97.10 \\
$0.75$ & 96.40 & 98.10 & 97.90 \\
$0.90$ & 92.40 & 95.50 & 96.10 \\
\bottomrule
\end{tabular}
\par\smallskip
\begin{minipage}{0.88\textwidth}
\footnotesize Notes: Coverage is reported in percent at $w=(0.5,0.5)^{\intercal}$ for
the Gaussian designs. Each entry equals 100 times the number of intervals covering the true quantile divided by the total number of replications. The regression bandwidth is
$\widehat h_{\rm cv}n^{-1/20}$.
\end{minipage}
\end{table}

\smallskip

In the degenerate design of Table~\ref{tab:dyadic-cover-all}, coverage ranges from $92.40\%$ to $98.10\%$ across all sample sizes and quantiles. At $n=60$, it is close to the nominal level across all five quantiles, ranging from $96.10\%$ to $97.90\%$. At the two outer quantiles, coverage increases from $93.40\%$ and $92.40\%$ at $n=15$ to $96.70\%$ and $96.10\%$ at $n=60$. The central quantiles exhibit moderate overcoverage, most noticeably at $n=30$, where coverage ranges from $97.30\%$ to $98.10\%$. Averaging over the five quantiles gives coverage of $95.28\%$, $96.98\%$, and $96.84\%$ as the number of nodes increases from $15$ to $30$ and $60$, respectively.

\smallskip

In the non-degenerate design, the coverage results improve as the sample size increases. Coverage averaged over the five quantiles increases from $87.76\%$ to $89.24\%$ and $90.24\%$ across the three sample sizes, and coverage is nondecreasing in $n$ at each individual quantile. At $\tau=0.90$, coverage rises from $83.40\%$ at $n=15$ to $87.50\%$ at $n=30$ and $87.70\%$ at $n=60$. At $\tau=0.10$, the corresponding values are $85.90\%$, $87.40\%$, and $88.90\%$. Coverage at the median is higher, at $91.90\%$, $92.10\%$, and $92.10\%$. The gains at the outer quantiles complement the improvements in estimation accuracy reported in Section~\ref{sec4.3} and highlight the benefit of larger networks for inference on the tails of the conditional distribution.

\smallskip

Taken together, the results show that the plug-in interval procedure with undersmoothing delivers coverage close to the nominal level in the degenerate design. In the non-degenerate design, average coverage moves toward the nominal level as the number of nodes increases from $15$ to $60$, with nondecreasing coverage at every quantile considered. Undersmoothing reduces the leading smoothing bias; accurate finite-sample inference also requires estimation of the variance components and an adequate normal approximation.

%%%%%%%%%%%%%%%%%%%

\section{An empirical application}\label{sec5}
\renewcommand{\theequation}{5.\arabic{equation}}
\setcounter{equation}{0}

This section illustrates the local linear dyadic quantile estimator on bilateral portfolio holdings drawn from the IMF Coordinated Portfolio Investment Survey (CPIS) for calendar year 2023, conditioned on country-level GDP from the World Bank's World Development Indicators (WDI). The estimation sample covers $n=51$ economies and $2038$ ordered dyads, and both the outcome and the conditioning variable enter on their natural monetary scales.

\subsection{Data description}\label{sec5.1}

The CPIS is the IMF's annual survey of cross-border portfolio holdings.\footnote{\url{https://data.imf.org/en/datasets/IMF.STA:PIP}.} For each pair of economies, it records the value of portfolio securities, that is, equity shares and debt securities, that residents of one economy hold in issuers located in the other, measured at year-end and reported in U.S. dollars. These bilateral positions are the most widely used measure of the geography of cross-border financial integration: they enter the international investment position accounts that countries publish under the international standard set out in the IMF's Balance of Payments and International Investment Position Manual, and they are a standard input to monitoring of global financial stability and to research on international portfolio allocation.

\smallskip

The unit of observation is the ordered dyad $(i,j)$, where $i$ is the reporting economy and $j$ is the counterpart economy, and the economic object is the year-end stock of portfolio assets
held by residents of $i$ in issuers located in $j$. The country-level covariate is nominal 2023 GDP in current U.S. dollars, taken from the World Bank's WDI \footnote{\url{https://data.worldbank.org/indicator/NY.GDP.MKTP.CD}.} and merged twice into the dyadic sample, once on the reporter side and once on the counterpart side, so that no bilateral controls are used.

\smallskip

The sample is restricted to the $n=51$ major economies, namely those whose 2023 nominal GDP is at least $\$100$ billion. These economies generate $n(n-1)=2550$ candidate ordered dyads. Among these dyads, $388$ positions are unavailable in the CPIS extract and $124$ are reported as zero. We restrict estimation to the remaining $2038$ positive positions, so the empirical quantiles describe recorded positive holdings conditional on reporter and counterpart GDP. The list of selected economies, ordered by 2023 GDP, is reported in Table~\ref{tab:selected-countries}.

\begin{table}[!htbp]
\centering
\small
\caption{Economies (with 2023 GDP $\geq\$100$ billion) in the annual CPIS sample.}
\label{tab:selected-countries}
\begin{tabular}{clclcl}
\toprule
Rank & Economy & Rank & Economy & Rank & Economy \\
\midrule
1  & United States       & 18 & Netherlands      & 35 & South Africa      \\
2  & China               & 19 & Switzerland      & 36 & Hong Kong, China  \\
3  & Germany             & 20 & Poland           & 37 & Colombia          \\
4  & Japan               & 21 & Belgium          & 38 & Romania           \\
5  & India               & 22 & Argentina        & 39 & Czech Republic    \\
6  & United Kingdom      & 23 & Sweden           & 40 & Pakistan          \\
7  & France              & 24 & Ireland          & 41 & Chile             \\
8  & Italy               & 25 & Austria          & 42 & Finland           \\
9  & Brazil              & 26 & Thailand         & 43 & Portugal          \\
10 & Canada              & 27 & Israel           & 44 & Kazakhstan        \\
11 & Korea, Republic of  & 28 & Singapore        & 45 & New Zealand       \\
12 & Mexico              & 29 & Norway           & 46 & Greece            \\
13 & Australia           & 30 & Bangladesh       & 47 & Hungary           \\
14 & Spain               & 31 & Philippines      & 48 & Ukraine           \\
15 & Indonesia           & 32 & Denmark          & 49 & Kuwait            \\
16 & Saudi Arabia        & 33 & Malaysia         & 50 & Slovakia          \\
17 & T\"urkiye           & 34 & Egypt            & 51 & Bulgaria          \\
\bottomrule
\end{tabular}
\end{table}

The portfolio outcome and both covariates refer to the same calendar year, so the specification is a clean cross section with contemporaneous node-level conditioning. Let $\mathrm{CPIS}_{ij}$
denote the bilateral asset position from reporter $i$ to counterpart $j$ and $\mathrm{GDP}_i$ denote nominal 2023 GDP of economy $i$. To keep both variables on legible numerical scales, the
estimation works with
\[
Y_{ij}=\mathrm{CPIS}_{ij}\text{ in USD billions},
\qquad
X_i=\mathrm{GDP}_i\text{ in USD trillions},
\]
so that the conditioning vector for the dyad is $W_{ij}=(X_i,X_j)^{^\intercal}\in\mathscr{R}^{2}$ and the evaluation point is $w=(x_\circ,x_\bullet)^{^\intercal}$.

\smallskip

Table~\ref{tab:summstats} reports descriptive statistics for the bilateral CPIS position on the USD-billion scale and for country-level GDP on the USD-trillion scale. Both variables have a heavy right tail typical of cross-border financial data: the median bilateral position is $\$0.56$ billion against a mean of $\$22.86$ billion and a maximum of nearly $\$2$ trillion, and country-level GDP ranges over more than two orders of magnitude with a mean of $\$1.89$ trillion against a median of $\$0.52$ trillion.

\begin{table}[!htbp]
\centering
\small
\caption{Descriptive statistics of the 2023 CPIS estimation sample.}
\label{tab:summstats}
\begin{tabular}{p{1.5cm}ccccccc}
\toprule
Variable & $n$ & Min & P25 & Median & Mean & P75 & Max \\
\midrule
$Y_{ij}$ & $2038$ & $0.000$ & $0.039$ & $0.556$ & $22.858$ & $5.671$ & $1984.535$ \\
$X_i$  & $51$ & $0.102$ & $0.346$ & $0.516$ & $1.890$ & $1.677$ & $27.292$ \\
\bottomrule
\end{tabular}
\medskip
\begin{minipage}{0.95\textwidth}
\footnotesize Notes: The bilateral CPIS ($Y_{ij}$) row summarizes the $2038$ ordered dyads in the estimation
sample of recorded positive positions. The GDP ($X_i$) row summarizes the $51$ selected economies at the
country level.
\end{minipage}
\end{table}

\subsection{Empirical results}\label{sec5.2}

We estimate the conditional $\tau$-quantile ${\sf Q}_\tau(w)$ of $Y_{ij}$ given $W_{ij}=w$ at $\tau\in\{0.10,0.25,0.50,0.75,0.90\}$ using the local linear dyadic quantile estimator (\ref{eq2.4})--(\ref{eq2.5}) with the product Gaussian kernel. The bandwidth $h$ is selected separately for each $\tau$ using the LTNO cross-validation criterion in Section~\ref{sec4.2}.

\smallskip

\begin{figure}[!htbp]
\centering
\includegraphics[width=0.85\textwidth]{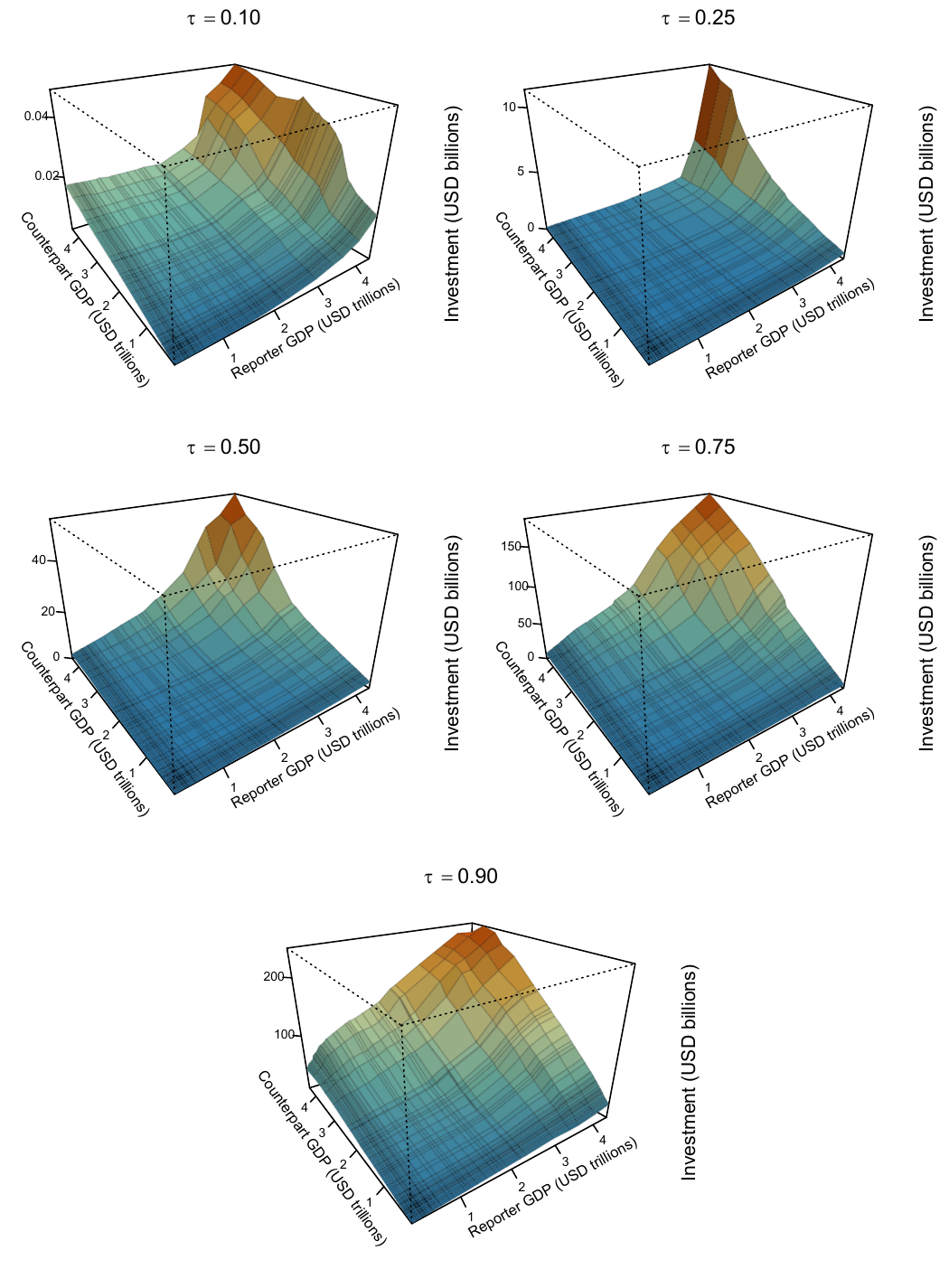}
\caption{The estimated conditional quantile surfaces $\widehat{\sf Q}_\tau(x_\circ,x_\bullet)$ of the
bilateral CPIS position (USD billions), as a function of reporter GDP $x_\circ$ and counterpart
GDP $x_\bullet$ (USD trillions), one panel per $\tau\in\{0.10,0.25,0.50,0.75,0.90\}$. Each panel
has its own vertical scale.}
\label{fig:quantile-surfaces}
\end{figure}

Figure~\ref{fig:quantile-surfaces} displays the estimated conditional quantile surfaces $\widehat{\sf Q}_\tau(x_\circ,x_\bullet)$ at each of the five quantile levels. The qualitative shape is the same across $\tau$: essentially flat in the lower-left quadrant where both economies are small, and rising convexly toward the upper-right corner where both are large. The vertical scale changes substantially across quantiles -- from sub-billion-dollar magnitudes at $\tau=0.10$ to the hundreds of billions at $\tau=0.90$ -- but the convexity persists at every quantile, indicating a multiplicative-style response to country size that a linear-in-levels specification would miss.

\begin{figure}[!htbp]
\centering
\includegraphics[width=0.78\textwidth]{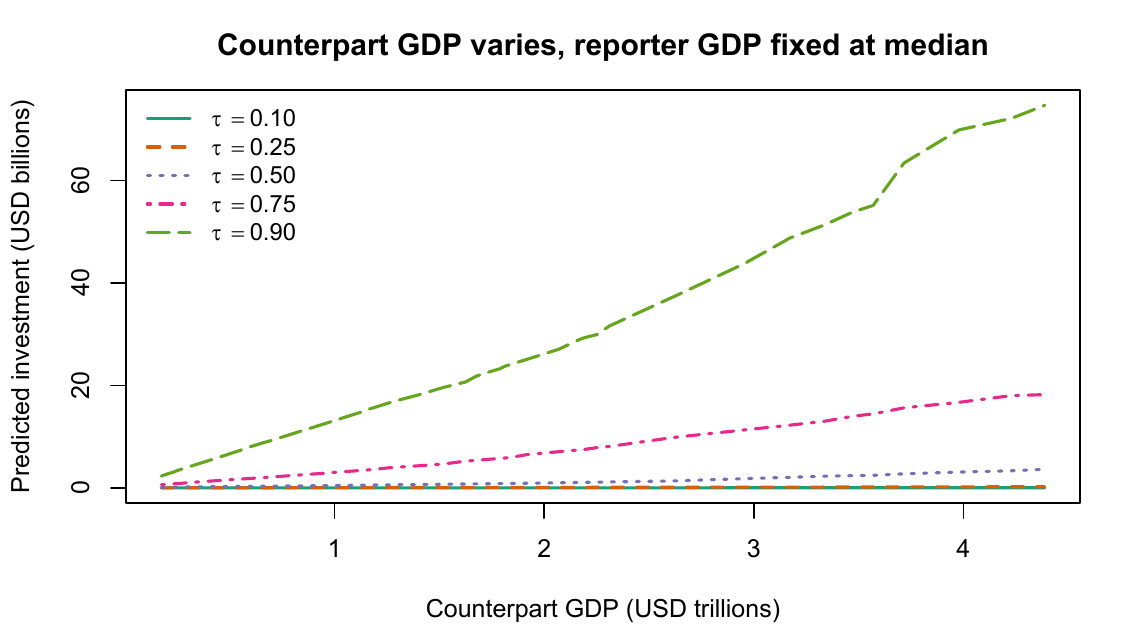}
\caption{The estimated conditional quantiles of the bilateral CPIS position $Y_{ij}$ (USD billions) as counterpart GDP varies, with reporter GDP fixed at the country-level GDP median. Curves
correspond to $\tau\in\{0.10,0.25,0.50,0.75,0.90\}$.}
\label{fig:slice-counterpart}
\end{figure}

\begin{figure}[!htbp]
\centering
\includegraphics[width=0.78\textwidth]{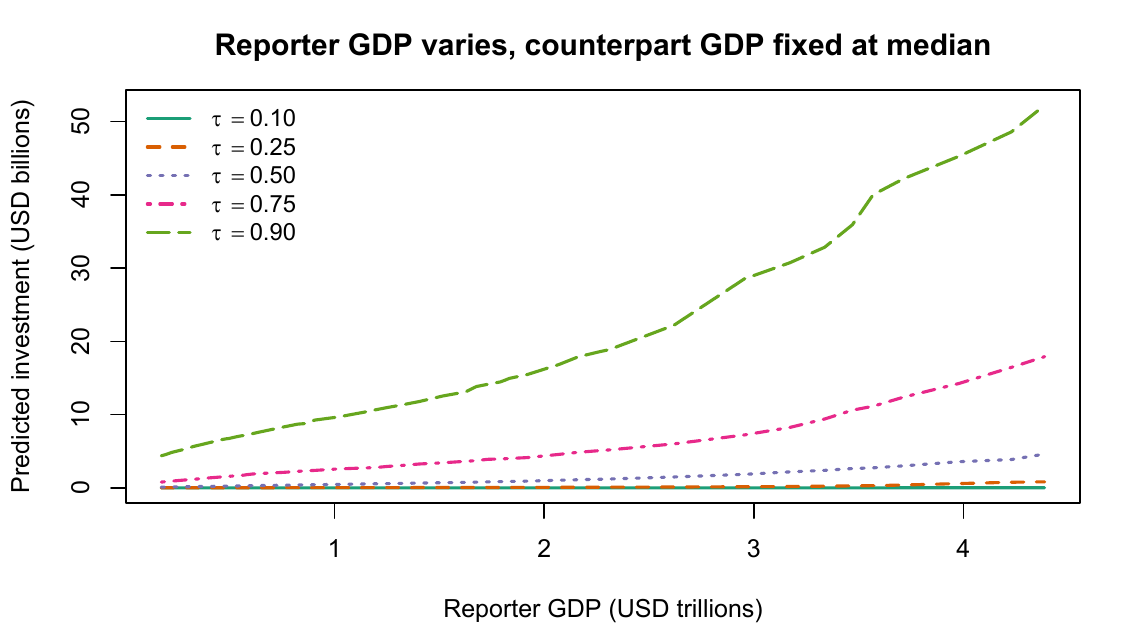}
\caption{The estimated conditional quantiles of the bilateral CPIS position $Y_{ij}$ (USD billions) as reporter GDP varies, with counterpart GDP fixed at the country-level GDP median. Curves
correspond to $\tau\in\{0.10,0.25,0.50,0.75,0.90\}$.}
\label{fig:slice-reporter}
\end{figure}

Figures~\ref{fig:slice-counterpart} and~\ref{fig:slice-reporter} offer a closer look at the surfaces along two median-fixed cross sections. In Figure~\ref{fig:slice-counterpart}, reporter GDP is held at the country-level GDP median and counterpart GDP varies between the $5$th and the $95$th percentile of the country-level GDP distribution; in Figure~\ref{fig:slice-reporter} the roles are reversed. The fitted curves shift upward in $\tau$ at every covariate value, consistent with the estimated quantile process being monotone in $\tau$, and they fan out sharply in the upper tail. In Figure~\ref{fig:slice-counterpart}, at $\tau=0.90$ the predicted bilateral position rises from roughly $\$2$ billion at the low end of the slice to roughly $\$75$ billion at the high end, with a clear convex curvature in the size of the partner economy; at $\tau\in\{0.10,0.25\}$ the curves remain essentially flat on the scale of the figure.

As a numerical supplement, Table~\ref{tab:lmh} evaluates $\widehat{\sf Q}_\tau(w)$ at the nine combinations of low, medium, and high covariate values, defined as the $20$th, $50$th, and $80$th
percentiles of the country-level GDP distribution. These percentiles correspond to GDPs of $\$0.34$, $\$0.52$, and $\$1.85$ trillion, respectively. Holding counterpart GDP at the median, raising reporter GDP from the low to the high value multiplies the estimated conditional quantile by a factor of $1.4$ at $\tau=0.10$, $3.7$ at $\tau=0.25$, $5.1$ at $\tau=0.50$, $3.4$ at $\tau=0.75$, and $2.6$ at $\tau=0.90$. The symmetric exercise that raises counterpart GDP while holding reporter GDP at the median yields multiplicative effects of $6.8$, $7.6$, $4.8$, $5.4$, and $5.4$ at the same five quantiles. Both sets of effects are sharply positive at every $\tau$, but their magnitudes differ across quantiles in a way that a single-equation conditional-mean specification cannot reveal: the reporter-side ratio is largest at $\tau=0.50$, whereas the counterpart-side ratio is largest at the lower-tail quantiles.

\begin{table}[!htbp]
\centering
\small
\caption{The estimated conditional quantiles of the bilateral CPIS position $Y_{ij}$ (USD billions) at selected GDP percentiles.}
\label{tab:lmh}
\begin{tabular}{llrrrrr}
\toprule
Reporter GDP & Counterpart GDP & $\tau=0.10$ & $\tau=0.25$ & $\tau=0.50$ & $\tau=0.75$ & $\tau=0.90$ \\
\midrule
Low    & Low    & $0.0010$ & $0.0082$ & $0.108$ & $0.724$ & $3.412$  \\
Low    & Medium & $0.0017$ & $0.0169$ & $0.174$ & $1.194$ & $5.716$  \\
Low    & High   & $0.0069$ & $0.0831$ & $0.734$ & $5.072$ & $22.162$ \\
Medium & Low    & $0.0010$ & $0.0118$ & $0.183$ & $1.120$ & $4.504$  \\
Medium & Medium & $0.0017$ & $0.0208$ & $0.253$ & $1.621$ & $6.875$  \\
Medium & High   & $0.0068$ & $0.0895$ & $0.868$ & $6.068$ & $24.158$ \\
High   & Low    & $0.0018$ & $0.0491$ & $0.743$ & $3.062$ & $12.000$ \\
High   & Medium & $0.0024$ & $0.0631$ & $0.884$ & $4.027$ & $15.009$ \\
High   & High   & $0.0071$ & $0.1682$ & $2.190$ & $13.860$ & $40.729$ \\
\bottomrule
\end{tabular}
\medskip
\begin{minipage}{0.95\textwidth}
\footnotesize Notes: Low, Medium, and High denote the $20$th, $50$th, and $80$th percentiles of the country-level GDP distribution, equal to $\$0.34$, $\$0.52$, and $\$1.85$ trillion
respectively. Each entry is an estimated conditional quantile $\widehat{\sf Q}_\tau(w)$ of bilateral portfolio holdings from (\ref{eq2.5}), expressed in USD billions. The evaluation points $x_\circ$ and $x_\bullet$ are reporter and counterpart GDP, respectively, expressed in USD trillions.
\end{minipage}
\end{table}

Our empirical application is intentionally parsimonious -- only reporter and counterpart GDP, on their natural USD scale, enter the conditioning vector. The surfaces, slices, and Table~\ref{tab:lmh} together expose distributional heterogeneity that a conditional-mean regression would miss: the marginal effect of either GDP differs across quantiles in both magnitude and curvature, and the
reporter--counterpart asymmetry is itself quantile-specific. The local linear quantile estimator captures the convex curvature and heavy right tail directly, sparing the user a logarithmic transformation that would otherwise be needed to linearize the relationship and that would carry its own misspecification risk.

%%%%%%%%%%%%%%%%%%%

\section{Conclusion}\label{sec6}

This paper studies the local linear nonparametric estimation of dyadic quantile regression within the conditionally independent dyad model framework which implies the properties of exchangeability and dyadic dependence. We establish the consistency and rate-adaptive standardized limit distribution theory for the proposed dyadic quantile regression estimation. The asymptotic theory shows that the dyadic quantile regression estimation convergence depends on whether the degeneracy exists for the underlying network generating process (in which case the dependence across dyads vanishes). This is substantially different from the standard limit theory of nonparametric quantile regression estimation for independent cross-sectional data or weakly dependent time series data. Furthermore, we consider estimating the asymptotic variance components which facilitates subsequent statistical inference. The finite-sample numerical performance of the proposed methodology is assessed via Monte-Carlo simulation. The empirical application to bilateral portfolio holdings (conditioned on country-level GDP) captures the convex curvature for quantile regression, allowing for  heavy right tails in the empirical data and exposing a quantile-specific reporter-counterpart asymmetry that the conditional mean regression would overlook.

%%%%%%%%%%%%%%%%%%%

\section*{Supplementary material}

The supplementary material contains the proofs of main results (in Appendix A), technical lemmas with proofs (in Appendix B) and extra comments to address practical issues in the variance estimation (in Appendix C).

\section*{Declaration of generative AI use}

OpenAI GPT-6 Astra assisted with proofreading and code generation to improve language and coding efficiency. The authors verified the content's originality and accuracy, including references, checked the tool's terms for publication suitability, and take full responsibility for the paper.

%%%%%%%%%%%%%%%%%%%%%%%%%%

\newpage

%%%%%%%%%%%%%%%%%%

\begin{center}
{\LARGE Supplementary Material for ``Nonparametric Dyadic Quantile Regression''}

\bigskip

{\small \textsc{Francesco Bravo}$^{*}$, \textsc{Degui Li }$^{\dag}$, \textsc{Zheng Li}$^{\ddag}$ }

{\small \medskip }

{\small\em $^{*}$University of York, $^{\dag}$ University of Macau, $^{\ddag}$ North Carolina State University}

\end{center}

\bigskip

\noindent Appendix A provides the detailed proofs of the main asymptotic results; Appendix B contains technical lemmas with proofs; and Appendix C discusses practical issues in the variance estimation. 

\section*{Appendix A:\ Proofs of the asymptotic results}
\renewcommand{\theequation}{A.\arabic{equation}}
\setcounter{equation}{0}

We start with some notation. Recall that $R_{ij}=Y_{ij}-{\sf Q}_{\tau}(W_{ij})$, $w=(x_\circ^{^\intercal}, x_\bullet^{^\intercal})^{^\intercal}$ and define 
\begin{eqnarray}
&&B_{ij}={\sf Q}_{\tau}(W_{ij})-{\sf Q}_{\tau}(w)-(W_{ij}-w)^{^\intercal}{\sf Q}_\tau^\prime(w),\notag\\
&&Z_{ij}=\left[1,\ (W_{ij}-w)^{^\intercal}/h\right]^{^\intercal}=\left[1,\ (X_{i}-x_\circ)^{^\intercal}/h, (X_{j}-x_\bullet)^{^\intercal}/h\right]^{^\intercal},\notag\\
&&\phi:=\phi(a,b)=\left[a-{\sf Q}_{\tau}(w),\ h(b-{\sf Q}_\tau^\prime(w))^{^\intercal}\right]^{^\intercal}.\notag
\end{eqnarray}
Note that the dyadic quantile regression loss function can be rewritten as 
\[
L_{\tau}(a,b)= \frac{1}{n\left(n-1\right)}\sum_{1\leq i \neq j\leq n}\rho_{\tau}\left(R_{ij}+B_{ij}-Z_{ij}^{^\intercal}\phi\right)K_{h}\left(W_{ij}-w\right).
\]
Throughout the proofs, we let $C$ denote a generic positive constant whose value may change from line to line.

\begin{proof}[{\bf Proof of Theorem \ref{Th1}}]\ (i) Let 
\[
\overline{L}_{\tau}(\phi) = \frac{1}{n(n-1)}\sum_{1\leq i \neq j\leq n}\left[\rho_{\tau}\left(R_{ij}+B_{ij}-Z_{ij}^{^\intercal}\phi\right)-\rho_{\tau}\left(R_{ij}+B_{ij}\right)\right]K_{h}(W_{ij}-w).
\]
Note that minimizing $L_{\tau}(a,b)$ with respect to $a$ and $b$ is equivalent to minimizing $\overline{L}_{\tau}(\phi)$ with respect to $\phi$. By \cite{Knight98}'s identity:
\[\rho_\tau(x-y)-\rho_\tau(x)=y\left[I(x\leq0)-\tau\right]+\int_0^y\left[I(x\leq z)-I(x\leq 0)\right]dz,\]
we have
\begin{eqnarray}
\overline{L}_{\tau}(\phi)&=&\phi^{^\intercal}\left\{\frac{1}{n(n-1)}\sum_{1\leq i \neq j\leq n}Z_{ij}\left[I\left(R_{ij}+B_{ij} \leq 0 \right)-\tau\right]K_{h}(W_{ij}-w)\right\}+\notag \\ 
&&\frac{1}{n(n-1)}\sum_{1\leq i \neq j\leq n}K_{h}(W_{ij}-w)\int_{0}^{Z_{ij}^{^\intercal}\phi}\left[I\left(R_{ij}+B_{ij} \leq u\right)- I\left(R_{ij}+B_{ij} \leq 0 \right)\right]du \notag\\
&=:&\phi^{^\intercal}\Pi_{1}+\Pi_{2}.\notag
\end{eqnarray}

For $\Pi_1$, consider the following decomposition:
\begin{eqnarray}
\Pi_1&=&\frac{1}{n(n-1)}\sum_{1\leq i \neq j\leq n}\psi_{\tau}\left(R_{ij}\right)Z_{ij}K_{h}\left(W_{ij}-w\right)+\notag \\
&&\frac{1}{n(n-1)}\sum_{1\leq i \neq j\leq n}\left[I\left(R_{ij}+B_{ij} \leq 0\right)-I\left(R_{ij} \leq 0\right)\right]Z_{ij}K_{h}\left(W_{ij}-w\right) \notag \\ 
&=:&\Pi_{1,1}+\Pi_{1,2},\label{eqA.1}
\end{eqnarray}
where $\psi_{\tau}(z)=I(z \leq 0) -\tau$. It follows from (\ref{eq2.3}) that ${\sf E}[\psi_{\tau}(R_{ij})|W_{ij}]=0$ {\em a.s.}, indicating that ${\sf E}[\Pi_{1,1}]=0$. This, together with Lemma \ref{le:1} in Appendix B, leads to 
\begin{equation}\label{eqA.2}
\Pi_{1,1}=O_P\left((nh^{d_{X}})^{-1/2}\right)(1+o(1))=o_P(1),
\end{equation}
whereas Lemma \ref{le:2} in Appendix B shows that
\begin{eqnarray}
\Pi_{1,2}&=&-\frac{1}{2}f_{Y|W}({\sf Q}_{\tau}(w)|w)f_W(w)h^2(1+o_P(1))\int s^{^\intercal} \frac{\partial^{2}{\sf Q}_{\tau}(w)}{\partial w \partial w^{^\intercal}}s \left(\begin{array}{c}
1\\
s
\end{array}\right) K(s)ds+o_P\left((nh^{d_{X}})^{-1/2}\right)\notag\\
&=&O_P\left(h^2\right)+o_P\left((nh^{d_{X}})^{-1/2}\right).\label{eqA.3}
\end{eqnarray}

By (\ref{eqB.rate}) in Lemma \ref{le:5}(ii), we have
\begin{equation}\label{eqA.4}
\widehat{\sf Q}_\tau(w)-{\sf Q}_\tau(w)=O_P\left(h^2+(nh^{d_X})^{-1/2}\right)=o_P(1).
\end{equation}
This proves (i).

\smallskip

(ii) Letting ${\boldsymbol\Lambda}(K)={\sf diag}\{1,\ \int ss^{^\intercal}K(s)ds\}$ and ${\bf V}=f_{Y|W}({\sf Q}_{\tau}(w)|w)f_{W}(w){\boldsymbol\Lambda}(K)$, it follows from Lemma \ref{le:5} in Appendix B that
\begin{equation}\label{eqA.5}
\widehat{\sf Q}_\tau(w)- {\sf Q}_\tau(w)=-e_1^{^\intercal}{\bf V}^{-1}\Pi_1+o_P(\varrho_n),\qquad \varrho_n=h^2+\sigma_n(w)
\end{equation}
where $e_1$ is a $(d_W+1)$-dimensional vector with the first element being one and the others being zeroes. Note that $\Pi_1=\Pi_{1,1}+\Pi_{1,2}$ as in (\ref{eqA.1}), and rewrite $\Pi_{1,1}$ as a $U$-statistic:
\[
\Pi_{1,1}=\frac{2}{n(n-1)}\sum_{1\leq i<j\leq n}T_{ij}(w)
\]
with
\[
T_{ij}(w)=\frac{1}{2}\left[Z_{ij}\psi_{\tau}(R_{ij})K_{h}(W_{ij}-w)+Z_{ji}\psi_{\tau}(R_{ji})K_{h}(W_{ji}-w)\right].
\]
Consider the following Hoeffding-type decomposition for $\Pi_{1,1}$:
\begin{eqnarray}
\Pi_{1,1}&=&\frac{2}{n}\sum_{i=1}^n\left\{{\sf E}[T_{ij}(w)|X_{i},U_{i}]-{\sf E}[T_{ij}(w)]\right\} +\notag\\ 
&&\frac{2}{n(n-1)}\sum_{1\leq i<j\leq n}\left\{T_{ij}(w)-{\sf E}[T_{ij}(w)|X_{i},X_{j},U_{i},U_{j}]\right\}+\notag\\  
&&\frac{2}{n(n-1)}\sum_{1\leq i<j\leq n}\left\{{\sf E}[T_{ij}(w)|X_{i},X_{j},U_{i},U_{j}]-{\sf E}[T_{ij}(w)|X_{i},U_{i}] -\right.\notag\\ 
&&\left.{\sf E}[T_{ji}(w)|X_{j},U_{j}]+{\sf E}[T_{ij}( w)]\right\}\notag\\ 
&=:&\Pi_{1,1,1}+\Pi_{1,1,2}+\Pi_{1,1,3}.\label{eqA.6}
\end{eqnarray}
The first term in the decomposition (\ref{eqA.6}) is a centered version of the H\'ajek projection for a classical U-statistic; the second term is a double array of conditionally independent random variables, and the third term is a degenerate second-order U-statistic. In addition, the three terms, $\Pi_{1,1,j}$, $j=1,2,3$, have mean zero and are pairwise orthogonal.

As the summands in $\Pi_{1,1,3}$ are uncorrelated, by the $c_r$ inequality and conditional Jensen's inequality, we have that 
\begin{eqnarray}
{\sf E}\left[\Vert \Pi_{1,1,3} \Vert ^2\right]&=&\frac{4}{[n(n-1)]^2} \sum_{1 \leq i < j \leq n}{\sf E}\left(\Vert {\sf E}[T_{ij}(w)|X_i,X_j,U_i,U_j ] - {\sf E}[T_{ij}(w)|X_i,U_i]-\right.\notag\\
&&\left.{\sf E}[T_{ji}(w)|X_j,U_j] + {\sf E}[T_{ij}(w)] \Vert ^2\right)\notag\\
&\leq& \frac{16}{n(n-1)}{\sf E}\left(\Vert {\sf E}[T_{12}(w)|X_1,X_2,U_1,U_2] \Vert ^2+2\Vert {\sf E}[T_{12}(w)|X_1,U_1] \Vert ^2 +\Vert {\sf E}[T_{12}(w)] \Vert^2\right) \notag \\
&\leq&\frac{64}{n(n-1)}{\sf E}\left(\Vert {\sf E}[T_{12}(w)|X_1,X_2,U_1,U_2]\Vert^2\right).\notag
\end{eqnarray}
In the case of non-degeneracy, i.e., $\nu_\tau(w)\neq 0$, we have that 
\[
{\sf E}\left[\Vert \Pi_{1,1,3} \Vert^2 \right]=O\left(\left(n^2h^{d_W}\right)^{-1}\right)=o\left(\left(nh^{d_X}\right)^{-1}\right),  
\]
indicating that ${\sf E}[\Vert \Pi_{1,1,3} \Vert^2 ]$ is of order smaller than $\sigma_n^2(w)$. In the degeneracy case when $\nu_\tau(w)=0$, since $R_{ij}$ are i.i.d. and independent of $\{X_i\}$, we readily have that ${\sf E}(T_{12}(w)|X_1,X_2,U_1,U_2)=0$, leading to the asymptotic negligibility of $\Pi_{1,1,3}$. Combining the above arguments, we obtain 
\begin{equation}\label{eqA.7}
{\sf E}\left[\Vert \Pi_{1,1,3} \Vert^2 \right]=o\left(\sigma_n^2(w)\right).
\end{equation}

It remains to establish the asymptotic distribution of $\theta^{^\intercal}(\Pi_{1,1,1}+\Pi_{1,1,2})$ for $\theta=-{\bf V}^{-1}e_1$. Following \cite{GNP24}'s technique, let $j(1)=2$ and $j(k)=1$ for $k\geq2$, and define
\begin{eqnarray}
A_{n,k}(w|\theta)&=&\frac{2\theta^{^\intercal}}{n}\left\{{\sf E}[T_{k,j(k)}(w)|X_{k},U_{k}]-{\sf E}[T_{k,j(k)}(w)]\right\},\quad k=1,\cdots,n,\notag\\
A_{n,n+k}(w|\theta)&=&\frac{2\theta^{^\intercal}}{n(n-1)}\left\{T_{1,k+1}(w)-{\sf E}[T_{1,k+1}(w)|X_{1},X_{k+1},U_{1},U_{k+1}]\right\},\notag\\
\vdots && \vdots\notag\\
A_{n,N(n)}(w|\theta)&=&\frac{2\theta^{^\intercal}}{n(n-1)}\left\{T_{n-1,n}(w)-{\sf E}[T_{n-1, n}(w)|X_{n-1},X_{n},U_{n-1},U_{n}]\right\},\notag
\end{eqnarray}
where $N(n)=n+n(n-1)/2$. Note that 
\[
\theta^{^\intercal}(\Pi_{1,1,1}+\Pi_{1,1,2})=\sum_{k=1}^{N(n)}A_{n,k}(w|\theta),
\]
and $\{A_{n,k}(w|\theta): 1\leq k\leq N(n)\}$ forms a triangular array of martingale differences due to the independence assumption on $\{X_i\}$, $\{U_i\}$ and $\{V_{ij}\}$. Hence, the martingale central limit theorem with a Lyapunov condition \citep[e.g.,][]{HH1980} can be adopted to prove the asymptotic normality for $\theta^{^\intercal}(\Pi_{1,1,1}+\Pi_{1,1,2})$. Let 
\[
\sigma_{n}^{2}(w|\theta)=\sum_{k=1}^{N(n)}{\sf Var}\left(A_{n,k}(w|\theta)\right).
\]
 Following the proof of Theorem 3.1 in \cite{GNP21}, we have, for $k=1,\cdots,n$,
\begin{eqnarray}
{\sf Var}\left(A_{n,k}(w|\theta)\right)&=&{\sf Var}\left(\frac{2\theta^{^\intercal}}{n}\left\{{\sf E}[T_{k,j(k)}(w)|X_{k},U_{k}]-{\sf E}[T_{k,j(k)}(w)]\right\}\right)\notag\\
&=&\frac{4}{n^2}{\sf Var}\left(\theta^{^\intercal}{\sf E}[T_{k,j(k)}(w)|X_{k},U_{k}]\right)\notag\\
&=&\frac{4}{n^2}\theta^{^\intercal}{\sf Cov}\left\{ T_{ik}(w), T_{jk}(w) \right\}\theta\label{eqA.8}
\end{eqnarray}
for pairwise distinct $i,j$ and $k$. The proof of Lemma \ref{le:1} in Appendix B shows that, if $\nu_\tau(w)\neq 0$, 
\begin{equation}\label{eqA.9}
\sum_{k=1}^n{\sf Var}\left(A_{n,k}(w|\theta)\right)\propto \left(nh^{d_X}\right)^{-1}, 
\end{equation}
whereas 
\begin{equation}\label{eqA.10}
{\sf Var}\left(A_{n,k}(w|\theta)\right)=O\left(\left(n^4h^{d_{W}}\right)^{-1}\right),\quad k=n+1,\cdots,N(n),
\end{equation}
so that 
\[
\sum_{k=n+1}^{N(n)}{\sf Var}\left(A_{n,k}(w|\theta)\right)= O\left(\left(n^2h^{d_W}\right)^{-1}\right).
\]

In the case of non-degeneracy, by (\ref{eqA.8})--(\ref{eqA.10}) and Assumption \ref{ass:3}(ii), we have 
\begin{equation}\label{eqA.11}
\sigma_{n}^{2}(w|\theta)=\sum_{k=1}^n{\sf Var}\left(A_{n,k}(w|\theta)\right)+o\left(\left(nh^{d_{X}}\right)^{-1}\right)\propto \left(nh^{d_X}\right)^{-1}.
\end{equation}
Hence, the asymptotic normality follows by verifying the following two conditions for Lyapunov's central limit theorem to hold, namely
\begin{eqnarray}
&&\sum_{k=1}^{N(n)}{\sf E}\left\vert \frac{A_{n,k}(w|\theta)}{\sigma_{n}(w|\theta)}\right\vert^\gamma\rightarrow0\ \ \text{for\ \ some\ \ } \gamma>2,\label{eqA.12}\\
and 
&&\sum_{k=1}^{N(n)}\left[\frac{A_{n,k}(w|\theta)}{\sigma_{n}(w|\theta)}\right]^2\stackrel{P}\longrightarrow1.\label{eqA.13}
\end{eqnarray}

Consider $\gamma=3$ in (\ref{eqA.12}). Notice that 
\begin{eqnarray}
{\sf E}\left[\lvert A_{n,k}(w|\theta)\rvert^{3}\right] &\leq& \frac{C}{\left[n(n-1)h^{d_W}\right]^{2}}{\sf E}\left[\lvert A_{n,k}(w|\theta)\rvert\right] \notag\\
&=&O\left(\left(n^{3}h^{d_{W}}\right)^{-2}\right),\label{eqA.14}
\end{eqnarray}
for $k=n+1,\cdots,N(n)$. As in the proof of Lemma \ref{le:1}, we have
\begin{equation}\label{eqA.15}
{\sf E}\left[\lvert A_{n,k}(w|\theta)\rvert^{3}\right]\leq \frac{C}{n^3}{\sf E}\left(\left\vert\theta^{^\intercal}{\sf E}[T_{k,j}(w)|X_{k},U_{k}]\right\vert^3\right)=O\left(\left(n^{3}h^{d_{W}}\right)^{-1}\right)
\end{equation}
for $k=1,\cdots,n$. Combining (\ref{eqA.11}), (\ref{eqA.14}) and (\ref{eqA.15}) and using the bandwidth restriction in Assumption \ref{ass:3}(ii), we have that
\[
\sum_{k=1}^{N(n)}{\sf E}\left\vert \frac{A_{n,k}(w|\theta)}{\sigma_{n}(w|\theta)}\right\vert^3=O\left(\left(nh^{d_{X}}\right)^{3/2}\left[\left(n^2h^{d_{W}}\right)^{-2}+\left(nh^{d_{X}}\right)^{-2}\right]\right)=o(1),
\]
indicating that (\ref{eqA.12}) is valid. 

We next turn to the stability condition (\ref{eqA.13}). Note that
\[
\frac{1}{\sigma_{n}^{2}(w|\theta)}\sum_{k=1}^{N(n)}\left(A_{n,k}^{2}(w|\theta)-{\sf E}\left[A_{n,k}^{2}(w|\theta)\right]\right)=\Xi_{n,1}+\Xi_{n,2}+\Xi_{n,3},
\]
where
\begin{eqnarray}
\Xi_{n,1}&=&\frac{4}{n^2\sigma_{n}^{2}(w|\theta)}\sum_{k=1}^n\left\{\left(\theta^{^\intercal}{\sf E}[T_{k,j(k)}(w)|X_{k},U_{k}]\right)^{2} - {\sf E}\left[\left(\theta^{^\intercal}{\sf E}[T_{k,j(k)}(w)|X_{k},U_{k}]\right)^{2}\right]\right\},\notag\\
\Xi_{n,2}&=&\frac{4}{[n(n-1)]^2\sigma_{n}^{2}(w|\theta)}\sum_{1\leq i<j\leq n}\left[\left(\theta^{^\intercal}\left\{T_{i,j}(w)-{\sf E}\left[T_{i,j}(w)|X_{i},X_{j},U_{i},U_{j}\right]\right\}\right)^2 \right.- \notag\\
&&{\sf E}\left. \left(\left(\theta^{^\intercal}\left\{T_{i,j}(w)-{\sf E}\left[T_{i,j}(w)|X_{i},X_{j},U_{i},U_{j}\right]\right\}\right)^2 |X_{i},X_{j},U_{i},U_{j}\right)\right],\notag\\
\Xi_{n,3}&=&\frac{4}{[n(n-1)]^2\sigma_{n}^{2}(w|\theta)}\sum_{1\leq i<j\leq n}\left[{\sf E}\left(\left(\theta^{^\intercal}\left\{T_{i,j}(w)-{\sf E}\left[T_{i,j}(w)|X_{i},X_{j},U_{i},U_{j}\right]\right\}\right)^2 |X_{i},X_{j},U_{i},U_{j}\right)\right.-\notag\\
&&\left.{\sf E}\left(\theta^{^\intercal}\left\{T_{i,j}(w)-{\sf E}\left[T_{i,j}(w)|X_{i},X_{j},U_{i},U_{j}\right]\right\}\right)^2\right].\notag
\end{eqnarray}
It is sufficient to show that 
\begin{equation}\label{eqA.16}
\Xi_{n,l}=o_P(1),\quad l=1,2,3,
\end{equation}
under (\ref{eqA.11}). Due to the independence assumption on $\{X_i\}$ and $\{U_i\}$, the summands in $\Xi_{n,1}$ are uncorrelated. Furthermore, the second moment of these summands is of order $h^{-3d_X}$. This, together with (\ref{eqA.11}), leads to 
\[
{\sf E}\left[\Xi_{n,1}^2\right]\leq \frac{C}{n^4\sigma_{n}^{4}(w|\theta)}nh^{-3d_X}\leq C\left(nh^{d_X}\right)^{-1}=o(1),
\] 
verifying (\ref{eqA.16}) for $l=1$. Note that the summands in $\Xi_{n,2}$ are also uncorrelated and the second moment of these summands is of order $h^{-3d_W}$, indicating that 
\[
{\sf E}\left[\Xi_{n,2}^2\right]\leq \frac{C}{n^8\sigma_{n}^{4}(w|\theta)}n^2h^{-3d_W}\leq C\left(nh^{d_X}\right)^{-4}=o(1),
\] 
verifying (\ref{eqA.16}) for $l=2$. It follows from Lemma \ref{le:4} that (\ref{eqA.16}) is satisfied for $l=3$. Then the stability condition (\ref{eqA.13}) is verified.

In the case of degeneracy, as $\{R_{ij}\}$ is i.i.d. and independent of $\{X_i\}$ and $\nu_\tau(w)=0$, following some standard calculation, we may show that 
\begin{eqnarray}
\sigma_{n}^{2}(w|\theta)&=&\sum_{k=1}^n{\sf Var}\left(A_{n,k}(w|\theta)\right)+\sum_{k=n+1}^{N(n)}{\sf Var}\left(A_{n,k}(w|\theta)\right)\notag\\
&\sim&\frac{1}{n(n-1)h^{d_W}}\tau(1-\tau)f_W(w)\theta^{^\intercal}{\boldsymbol\Omega} (K)\theta,\qquad \label{eqA.17}
\end{eqnarray}
where ${\boldsymbol\Omega}(K)=\int (1,s^{^\intercal})^{^\intercal}(1,s^{^\intercal})K^2(s)ds$. Lyapunov's central limit theorem still applies to this case. It remains to verify (\ref{eqA.12}) and (\ref{eqA.13}), which we sketch as follows. To verify (\ref{eqA.12}), we note that 
\[
\sigma_n^3(w|\theta)\propto \left(n^{2}h^{d_W}\right)^{-3/2}\quad \text{and}\quad {\sf E} \vert A_{n,k}(w,\theta) \vert^3=O\left(\left(n^{3}h^{d_{W}}\right)^{-2}\right).
\]
Hence, we have
\[
\sum_{k=n+1}^{N(n)}{\sf E}\left\vert \frac{A_{n,k}(w|\theta)}{\sigma_n(w|\theta)} \right\vert ^3=O\left(\frac{n^2}{\left(n^{3}h^{d_{W}}\right)^{2}}\left(n^2h^{d_W}\right)^{3/2}\right)=O\left(\frac{1}{nh^{d_X}}\right)=o(1),
\]
proving (\ref{eqA.12}) with $\gamma=3$. To verify (\ref{eqA.13}), write $A_{n,ij}(w|\theta)$ for the increment associated with the unordered pair $i<j$. Under case (b), the node increments vanish. The squares of increments on disjoint pairs are independent, but squares on pairs sharing a node need not be independent because of their kernel weights. Boundedness of the scores and kernel, together with a change of variables, gives
\[
{\sf E}[A_{n,ij}^4(w|\theta)]=O\left(n^{-8}h^{-3d_W}\right),\qquad
\left|{\sf Cov}\left(A_{n,ij}^2(w|\theta),A_{n,ik}^2(w|\theta)\right)\right|=O\left(n^{-8}h^{-5d_X}\right)
\]
for distinct $i,j,k$, with the same bound for the other shared-node configurations. There are $O(n^2)$ diagonal terms and $O(n^3)$ shared-node terms. Consequently,
\begin{eqnarray}
{\sf Var}\left(\sum_{k=n+1}^{N(n)}A_{n,k}^2(w|\theta)\right)
&=&O\left(n^{-6}h^{-3d_W}+n^{-5}h^{-5d_X}\right),\notag\\
\frac{1}{\sigma_n^4(w|\theta)}{\sf Var}\left(\sum_{k=n+1}^{N(n)}A_{n,k}^2(w|\theta)\right)
&=&O\left((nh^{d_X})^{-2}+(nh^{d_X})^{-1}\right)=o(1),\notag
\end{eqnarray}
where the second line uses (\ref{eqA.17}). Since the expectation of the sum of squared increments is $\sigma_n^2(w|\theta)$, Chebychev's inequality proves (\ref{eqA.13}).

Then, by Lyapunov's central limit theorem, we prove that
\begin{equation}\label{eqA.18}
\frac{1}{\sigma_{n}(w|\theta)}\sum_{k=1}^{N(n)}A_{n,k}(w|\theta) \overset{d}{\longrightarrow} {\sf N}(0,1)
\end{equation}
for both the non-degeneracy and degeneracy cases. Furthermore, by the block diagonal structure of ${\boldsymbol\Lambda}(K)$, we readily have that $e_1^{^\intercal}{\bf V}^{-1}=e_1^{^\intercal}/[f_{Y|W}({\sf Q}_{\tau}(w)|w)f_{W}(w)]$. Hence, by Lemma \ref{le:2}, $-e_1^{^\intercal}{\bf V}^{-1}\Pi_{1,2}=b_\tau(w)h^2+o_P(\varrho_n)$, which, together with (\ref{eqA.5}) and $h^2=O(\sigma_n(w))$, gives
\[
\widehat{\sf Q}_\tau(w)-{\sf Q}_\tau(w)-b_\tau(w)h^2=-e_1^{^\intercal}{\bf V}^{-1}\Pi_{1,1}+o_P(\sigma_n(w)).
\]
Setting $\theta^{^\intercal}=-e_1^{^\intercal}{\bf V}^{-1}$ and using (\ref{eqA.6}), (\ref{eqA.7}), (\ref{eqA.17}) and (\ref{eqA.18}), we complete the proof of Theorem \ref{Th1}(ii). 

\end{proof}

\begin{proof}[{\bf Proof of Proposition \ref{prop:1}}]

For notational simplicity, we let $\widehat{\sf Q}_{\mathscr {M}_1}=\widehat{\sf Q}_\tau^{\mathscr {M}_1}(w)$, which is independent of the sub-sample $\mathscr{M}_2$. Define an infeasible conditional density function estimation with sample-splitting:
\begin{equation}\label{eqA.19}
\check{f}_{Y|W}({\sf Q}_\tau(w)|w)=\frac{\sum_{n_1+1\leq i\neq j\leq n}k_{b_1}(Y_{ij}- {\sf Q}_\tau(w))H_{b_2}(W_{ij}-w)}{\sum_{n_1+1\leq i\neq j\leq n}H_{b_2}(W_{ij}-w)}.
\end{equation}
By the triangle inequality,
\begin{eqnarray}
\left\vert \widehat{f}_{Y|W}({\sf Q}_\tau(w)|w)-f_{Y|W}({\sf Q}_\tau(w)|w)\right\vert &\leq& \left\vert \widehat{f}_{Y|W}( {\sf Q}_\tau(w)|w)-\check{f}_{Y|W}({\sf Q}_\tau(w)|w)\right\vert + \notag\\
&&\left\vert \check{f}_{Y|W}({\sf Q}_\tau(w)|w)-f_{Y|W}({\sf Q}_\tau(w)|w)\right\vert.\label{eqA.20}
\end{eqnarray}

For the second term in $(\ref{eqA.20})$, we first consider the numerator. With a standard change of variable and Taylor's expansion, by Assumptions \ref{ass:2}, \ref{ass:4} and \ref{ass:5}, we may show that 
\begin{eqnarray}
&&{\sf E}\left[k_{b_{1}}(Y_{ij}-\textsf{Q}_\tau(w))H_{b_{2}}(W_{ij}-w)\right]\notag\\
&=& \frac{1}{b_{1}b_{2}^{d_W}}\int\int k\left(\frac{y-\textsf{Q}_\tau(w)}{b_{1}}\right)H\left(\frac{w^\prime-w}{b_{2}}\right)f_{Y|W}(y|w^\prime)f_W(w^\prime)dydw^\prime\nonumber \\  
&=&\int\int k(u)H(v)f_{Y|W}(\textsf{Q}_\tau(w)+ub_{1}|w+vb_{2})f_W(w+vb_2)dudv\notag\\
&=& f_{Y|W}(\textsf{Q}_\tau(w)|w)f_W(w)+o(1). \label{eqA.21}
\end{eqnarray}
We next turn to the variance calculation. Define
\[
D_{ij}=k_{b_1}(Y_{ij}-{\sf Q}_\tau(w))H_{b_2}(W_{ij}-w),\qquad D_{ij}^{\circ}=\frac{D_{ij}+D_{ji}}{2}.
\]
Letting $\varpi_{n}=(n-(n_1+1))(n-n_1)$ and using the dyadic dependence property, we have
\[
{\sf Var}\left(\frac{1}{\varpi_{n}}\sum_{n_1+1 \leq i \neq j \leq n}D_{ij}\right)=\Upsilon_{1}+\Upsilon_{2},
\]
where
{\small\begin{eqnarray}
\Upsilon_{1}&=&\frac{4}{\varpi_{n}^2}\sum_{n_1+1\leq i<j\leq n}{\sf Var}\left(D_{ij}^{\circ}\right),\notag \\
\Upsilon_{2}&=&\frac{4}{\varpi_{n}^2}\sum_{(i_1,j_1,i_2,j_2)\in{\cal D}_2}{\sf Cov}\left(D_{i_1j_1}^{\circ},D_{i_2j_2}^{\circ}\right).\notag
\end{eqnarray}}
Here ${\cal D}_2$ is the set of $(i_1,j_1,i_2,j_2)$ such that one of the following four scenarios holds: 

\begin{itemize}

\item $n_1+1\leq i_1<j_1\leq n,\ n_1+1\leq i_2<j_2\leq n,\ i_1=i_2,\ j_1\neq j_2$;

\item $n_1+1\leq i_1<j_1\leq n,\ n_1+1\leq i_2<j_2\leq n,\ j_1=j_2,\ i_1\neq i_2$;

\item $n_1+1\leq i_1<j_1\leq n,\ n_1+1\leq i_2<j_2\leq n,\ i_1= j_2$;

\item $n_1+1\leq i_1<j_1\leq n,\ n_1+1\leq i_2<j_2\leq n,\ i_2=j_1$.

\end{itemize}
By Assumption \ref{ass:2}(ii), ${\sf E}[k_{b_1}^2(Y_{ij}-{\sf Q}_\tau(w))\mid W_{ij}]\leq C/b_1$ for $W_{ij}$ in a neighborhood of $w$. Hence ${\sf E}(D_{ij}^2)=O((b_1b_2^{d_W})^{-1})$. For pairwise distinct $i,j,k$, the conditional Cauchy--Schwarz inequality gives
\[
{\sf E}\left[\left|k_{b_1}(Y_{ij}-{\sf Q}_\tau(w))k_{b_1}(Y_{ik}-{\sf Q}_\tau(w))\right|\,\middle|\,X_i,X_j,X_k\right]\leq\frac{C}{b_1}
\]
on the joint support of the covariate kernel weights. Integrating over the three independent covariate vectors and using the boundedness and compact support of $H$, we obtain ${\sf E}|D_{ij}D_{ik}|=O((b_1b_2^{d_X})^{-1})$. The same bound holds for the other shared-index configurations and consequently for the covariances of the symmetrized summands. Therefore, we have

\begin{equation}\label{eqA.22}
\Upsilon_{1}=O\left(\frac{1}{\varpi_n b_{1}b_{2}^{d_W}}\right)=O\left(\frac{1}{n^2b_{1}b_{2}^{d_W}}\right)=o(1)
\end{equation}
and 
\begin{equation}\label{eqA.23}
\Upsilon_{2}=O\left(\frac{n^3}{\varpi_n^2 b_1b_{2}^{d_X}}\right)=O\left(\frac{1}{nb_1b_{2}^{d_X}}\right)=o(1).
\end{equation}
The last equality follows from Assumption \ref{ass:5}: $(nh^{d_X})^{-1/2}=o(b_1)$ implies $\sqrt n\,b_1\to\infty$, whereas $nb_2^{d_W}\to\infty$ and $d_W=2d_X$ imply $\sqrt n\,b_2^{d_X}\to\infty$. Thus $nb_1b_2^{d_X}\to\infty$.
It is easy to verify that
\begin{equation}\label{eqA.24}
\varpi_n^{-1} \sum_{n_1+1 \leq i \neq j \leq n}H_{b_2}(W_{ij}-w)=f_W(w)+o_P(1).  
\end{equation}
Combining (\ref{eqA.21})--(\ref{eqA.24}), we readily have that 
\begin{equation}\label{eqA.25}
\left\vert \check{f}_{Y|W}({\sf Q}_\tau(w)|w)-f_{Y|W}({\sf Q}_\tau(w)|w)\right\vert=o_P(1).
\end{equation}

It remains to prove that 
\begin{equation}\label{eqA.26}
\left\vert \widehat{f}_{Y|W}( {\sf Q}_\tau(w)|w)-\check{f}_{Y|W}({\sf Q}_\tau(w)|w)\right\vert=o_P(1).
\end{equation}
Write 
$$
\widehat\delta:=\widehat\delta_{{\mathscr M}_1}=\widehat{\sf Q}_{{\mathscr M}_1}-{\sf Q}_\tau(w)\quad \text{and}\quad r_n=h^2+(nh^{d_X})^{-1/2}.
$$ 
Applying (\ref{eqB.rate}) in Lemma \ref{le:5} to the subsample $\mathscr{M}_1$ and using Assumption \ref{ass:5}, we obtain
\begin{equation}
\frac{|\widehat\delta|}{b_1}=O_P\Big(\frac{r_n}{b_1}\Big)=o_P(1). \label{eqA.27}
\end{equation}
For any $y\in\mathbb{R}$ and $x\in\mathbb{R}$, the fundamental theorem of calculus gives
\begin{equation}
k_{b_1}\big(y-{\sf Q}_\tau(w)-x\big)-k_{b_1}\big(y-{\sf Q}_\tau(w)\big)=-\frac{x}{b_1}\int_0^1 \dot k_{b_1}\big(y-{\sf Q}_\tau(w)-xt\big)\,dt, \label{eqA.28}
\end{equation}
where $\dot k_{b_1}(\cdot)=b_1^{-1}\dot k(\cdot/b_1)$. Applying (\ref{eqA.28}) with $y=Y_{ij}$ and $x=\widehat\delta$ to each summand, we obtain
\begin{equation}
\Big|\widehat f_{Y|W}({\sf Q}_\tau(w)|w)-\check f_{Y|W}({\sf Q}_\tau(w)|w)\Big|\leqslant \frac{|\widehat\delta|}{b_1}\cdot
\frac{S_n(\widehat\delta)}{\big|\varpi_n^{-1}\sum_{n_1+1\leqslant i\neq j\leqslant n}H_{b_2}(W_{ij}-w)\big|}, 
\label{eqA.29}
\end{equation}
where
\[
S_n(x)=\frac{1}{\varpi_n}\sum_{n_1+1\leqslant i\neq j\leqslant n}\int_0^1\big|\dot k_{b_1}\big(Y_{ij}-{\sf Q}_\tau(w)-xt\big)\big|\,dt\,\big|H_{b_2}(W_{ij}-w)\big|,\qquad x\in\mathbb{R}.
\]
For each fixed $x$, $S_n(x)$ is a nonnegative function of $\mathscr{M}_2$ alone, and its summands are identically distributed. Then, by Assumptions \ref{ass:2}(ii) and \ref{ass:4}, for any deterministic $x$ with $|x|\leqslant\epsilon$ and $\epsilon$ being a sufficiently small positive number, a change of variables as in (\ref{eqA.21}) yields
\begin{align}
\mathsf{E}[S_n(x)]
&=\int_0^1\!\!\int\!\!\int |\dot k(u)|\,|H(v)|\,
f_{Y|W}\big({\sf Q}_\tau(w)+xt+ub_1\,\big|\,w+vb_2\big)\,f_W(w+vb_2)\,du\,dv\,dt
\nonumber\\
&\leqslant C\int|\dot k(u)|\,du\int|H(v)|\,dv=:C_\circ<\infty, \label{eqA.30}
\end{align}
where $C_\circ$ does not depend on $x$ or $n$. Letting $\sigma(\mathscr{M}_1)$ denote the $\sigma$-field generated by ${\mathscr M}_1$ and 
\[
\mathcal{E}_n=\left\{|\widehat\delta|\leqslant\epsilon\right\}\in\sigma(\mathscr{M}_1),
\]
by (\ref{eqA.27}), $\mathsf{P}(\mathcal{E}_n)\to1$. Since $\widehat\delta$ is $\sigma(\mathscr{M}_1)$-measurable and $\mathscr{M}_2$ is independent of $\mathscr{M}_1$, it follows from (\ref{eqA.30}) that $\mathsf{E}[S_n(\widehat\delta)\,|\,\mathscr{M}_1]\leqslant C_\circ$ conditional on $\mathcal{E}_n$. Hence, for any $M>0$, the conditional Markov inequality gives
\[
\mathsf{P}\big(S_n(\widehat\delta)>M\big)\leqslant \mathsf{P}(\mathcal{E}_n^c)+\mathsf{E}\Big[I(\mathcal{E}_n)\, \mathsf{P}\big(S_n(\widehat\delta)>M\,\big|\, \mathscr{M}_1\big)\Big]\leqslant o(1)+\frac{C_\circ}{M},
\]
and thus $S_n(\widehat\delta)=O_P(1)$. Finally, by (\ref{eqA.24}) the denominator in (\ref{eqA.29}) converges in probability to $f_W(w)>0$. Combining these results with (\ref{eqA.27}), the right-hand side of (\ref{eqA.29}) is $o_P(1)\cdot O_P(1)\cdot O_P(1)=o_P(1)$, which proves (\ref{eqA.26}). Together with (\ref{eqA.25}) and (\ref{eqA.20}), this completes the proof of Proposition \ref{prop:1}. 

\end{proof}

%%%%%%%%%%%%%%%%%%

\section*{Appendix B:\ Technical lemmas with proofs}
\renewcommand{\theequation}{B.\arabic{equation}}
\setcounter{equation}{0}

This appendix contains some technical lemmas which have been used in the proofs of the main theoretical results in Appendix A.

\begin{lemma}\label{le:1}

Suppose that Assumptions \ref{ass:1}, \ref{ass:2}(i) and \ref{ass:3} are satisfied. Let $\Pi_{1,1}$ be defined as in (\ref{eqA.1}). Then we have ${\sf Var}(\Pi_{1,1})=O((nh^{d_{X}})^{-1})+O((n(n-1)h^{d_W})^{-1})$.

\end{lemma}

\begin{proof}[{\bf Proof of Lemma \ref{le:1}}]  Rewrite $\Pi_{1,1}$ as the following $U$-statistic:
\[
\Pi_{1,1}=\frac{2}{n(n-1)}\sum_{i<j}T_{ij}(w),
\]
where
\[
T_{ij}(w)=\frac{1}{2}\left[Z_{ij}\psi_{\tau}(R_{ij})K_{h}(W_{ij}-w)+Z_{ji}\psi_{\tau}(R_{ji})K_{h}(W_{ji}-w)\right].
\]
It follows from the dyadic dependence property that
\begin{equation}\label{eqB.1}
{\sf Var}(\Pi_{1,1})=\frac{4}{n^2(n-1)^2}\sum_{i<j}{\sf Var}(T_{ij}(w))+\frac{4}{n^2(n-1)^2}\sum_{(i_1,j_1,i_2,j_2)\in{\cal D}}{\sf Cov}(T_{i_1j_1}(w), T_{i_2j_2}(w)),
\end{equation}
where ${\cal D}$ is the set of $(i_1,j_1,i_2,j_2)$ such that one of the following four scenarios hold: 

\begin{itemize}

\item $i_1<j_1,\ i_2<j_2,\ i_1=i_2,\ j_1\neq j_2$;

\item $i_1<j_1,\ i_2<j_2,\ j_1=j_2,\ i_1\neq i_2$;

\item $i_1<j_1,\ i_2<j_2,\ i_1= j_2$;

\item $i_1<j_1,\ i_2<j_2,\ i_2=j_1$.

\end{itemize}

We next only derive the order for the $(1,1)$-entry of ${\sf Var}(T_{ij}(w))$ 
\[
{\sf Var}(T_{ij}^{\circ}(w))\ \ \text{with}\ \ T_{ij}^{\circ}(w)=\frac{1}{2}\left[\psi_{\tau}(R_{ij})K_{h}(W_{ij}-w)+\psi_{\tau}(R_{ji})K_{h}(W_{ji}-w)\right],
\]
as the order for the other entries is the same. Using Assumptions \ref{ass:1}, \ref{ass:2}(i) and \ref{ass:3}, and following a standard calculation for the kernel-weighted component, we may show that
\begin{eqnarray}
{\sf Var}(T_{ij}^{\circ}(w))&\leq&\frac{1}{2}{\sf Var}\left(\psi_\tau(R_{ij}) K_{h}(W_{ij}-w)\right)+\frac{1}{2}{\sf Var}\left(\psi_\tau(R_{ji}) K_{h}(W_{ji}-w)\right)\notag\\
&=&{\sf E}\left[K_{h}^2(W_{ij}-w) {\sf E}\left(\psi_\tau^2(R_{ij}) |X_{i},X_{j}\right)\right]\notag\\
&=&\frac{\tau(1-\tau)}{h^{d_{W}}} \int f_{W}(w+hs)K^2(s)ds\notag\\
&=&O\left(h^{-d_{W}}\right),\notag
\end{eqnarray}
indicating that
\begin{equation}\label{eqB.2}
\frac{1}{n^2(n-1)^2}\sum_{i<j}{\sf Var}(T_{ij}(w))=O\left(\left(n(n-1)h^{d_{W}}\right)^{-1}\right).
\end{equation}

We next derive the order for the second term on the right hand side of (\ref{eqB.1}), and consider the scenario $(i_1<j_1,\ i_2<j_2,\ i_1=i_2,\ j_1\neq j_2)$ in ${\cal D}$ for the $(1,1)$-entry of ${\sf Cov}(T_{i_1j_1}(w), T_{i_2j_2}(w))$. Without loss of generality, we let $i_1=i_2=1$, $j_1=2$ and $j_2=3$. Note that
\begin{eqnarray}
{\sf Cov}(T_{12}^{\circ}(w), T_{13}^{\circ}(w))&=&\frac{1}{4}\left[{\sf Cov}\left(\psi_{\tau}(R_{12})K_{h}(W_{12}-w),\ \psi_{\tau}(R_{13})K_{h}(W_{13}-w)\right)+\right.\notag\\
&&{\sf Cov}\left(\psi_{\tau}(R_{12})K_{h}(W_{12}-w),\ \psi_{\tau}(R_{31})K_{h}(W_{31}-w)\right)+\notag\\
&&{\sf Cov}\left(\psi_{\tau}(R_{21})K_{h}(W_{21}-w),\ \psi_{\tau}(R_{13})K_{h}(W_{13}-w)\right)+\notag\\
&&\left.{\sf Cov}\left(\psi_{\tau}(R_{21})K_{h}(W_{21}-w),\ \psi_{\tau}(R_{31})K_{h}(W_{31}-w)\right)\right].\notag
\end{eqnarray}
Let $V_\tau(x_1,x_2,x_3)={\sf E}\left(\psi_\tau(R_{12})\psi_\tau(R_{13}) |X_{1}=x_1,X_{2}=x_2, X_3=x_3\right)$ and $w=(x_\circ^{^\intercal}, x_\bullet^{^\intercal})^{^\intercal}$. Note that
\begin{eqnarray}
&&{\sf Cov}\left(\psi_{\tau}(R_{12})K_{h}(W_{12}-w),\ \psi_{\tau}(R_{13})K_{h}(W_{13}-w)\right)\notag\\
&=&{\sf E}\left[K_{h}(W_{12}-w) K_{h}(W_{13}-w) V_\tau(X_1,X_2,X_3)\right]\notag\\
&=&\frac{1}{h^{2d_{W}}} \int V_\tau(x_1,x_2,x_3)  K\left(\frac{x_{1}-x_{\circ}}{h},\frac{x_{2}-x_{\bullet}}{h}\right)K\left(\frac{x_{1}-x_{\circ}}{h},\frac{x_{3}-x_{\bullet}}{h}\right) f_{X}(x_1)f_{X}(x_2)f_{X}(x_3)dx_1dx_2dx_3\notag\\
&=&\frac{1}{h^{d_{X}}}V_\tau(x_\circ,x_\bullet,x_\bullet) f_{X}(x_\circ)f_{X}^2(x_\bullet)\Delta_\circ(K)=O\left(h^{-d_{X}}\right),\notag
\end{eqnarray}
where
\[
\Delta_\circ(K)=\int K\left(s_{\circ}, s_{\bullet}\right)K\left(s_{\circ}, s_{\bullet}^\prime\right)ds_\circ ds_\bullet ds_\bullet^\prime.
\]
We can derive the same order for the other covariances, and consequently obtain 
\begin{equation}\label{eqB.3}
{\sf Cov}(T_{12}^{\circ}(w), T_{13}^{\circ}(w))=O\left(h^{-d_{X}}\right).
\end{equation}
Furthermore, we may show that (\ref{eqB.3}) holds for any entry of ${\sf Cov}(T_{i_1j_1}(w), T_{i_2j_2}(w))$, $(i_1,j_1,i_2,j_2)\in{\cal D}$. Note that the cardinality of ${\cal D}$ is of order $n^3$. Hence, we have
\begin{equation}\label{eqB.4}
\frac{1}{n^2(n-1)^2}\sum_{(i_1,j_1,i_2,j_2)\in{\cal D}}{\sf Cov}(T_{i_1j_1}(w), T_{i_2j_2}(w))=O\left(\left(nh^{d_{X}}\right)^{-1}\right).
\end{equation} 
With (\ref{eqB.1}), (\ref{eqB.2}) and (\ref{eqB.4}), noting that $nh^{d_X}\rightarrow\infty$, we complete the proof of Lemma \ref{le:1}.

%{\red In the degenerate case (b) of Theorem \ref{Th1}(ii), the covariance terms in (\ref{eqB.1}) vanish. Conditional on ${\cal F}_X=\sigma\{X_1,\cdots,X_n\}$, the variables $\psi_\tau(R_{i_1j_1})$ and $\psi_\tau(R_{i_2j_2})$ are independent for distinct ordered pairs, with ${\sf E}[\psi_\tau(R_{ij})\,|\,{\cal F}_X]=0$. Hence
%\begin{equation}\label{eqB.deg}
%{\sf Var}(\Pi_{1,1})=O\Big(\big(n(n-1)h^{d_W}\big)^{-1}\Big)=O\big(\sigma_n^2(w)\big).
%\end{equation}
%Under non-degeneracy, $(nh^{d_X})^{-1}\asymp\sigma_n^2(w)$, so ${\sf Var}(\Pi_{1,1})=O(\sigma_n^2(w))$ in both cases.}

\end{proof}

\begin{lemma}\label{le:2}

Suppose that Assumptions \ref{ass:1}, \ref{ass:2} and \ref{ass:3}(i)(ii) are satisfied. Let $\Pi_{1,2}$ be defined as in (\ref{eqA.1}). Then we have 
\begin{equation}\label{eqB.5}
\Pi_{1,2}=-\frac{1}{2}f_{Y|W}({\sf Q}_{\tau}(w)|w)f_W(w)h^2(1+o_P(1))\int s^{^\intercal} \frac{\partial^{2}{\sf Q}_{\tau}(w)}{\partial w \partial w^{^\intercal}}s \left(\begin{array}{c}
1\\
s
\end{array}\right) K(s)ds+o_P\left((nh^{d_{X}})^{-1/2}\right).
\end{equation}
Furthermore, when either (a) or (b) of Theorem \ref{Th1}(ii) holds, then
\begin{equation}\label{eqB.5b}
\Pi_{1,2}=-\frac{1}{2}f_{Y|W}({\sf Q}_{\tau}(w)|w)f_W(w)h^2\int s^{^\intercal} \frac{\partial^{2}{\sf Q}_{\tau}(w)}{\partial w \partial w^{^\intercal}}s \left(\begin{array}{c}
1\\
s
\end{array}\right) K(s)ds+o_P\big(h^2+\sigma_n(w)\big).
\end{equation}

\end{lemma}

\begin{proof}[{\bf Proof of Lemma \ref{le:2}}] We first derive the mean of $\Pi_{1,2}$. By Assumptions \ref{ass:1}--\ref{ass:3} and Taylor's expansion, we readily have that
\begin{eqnarray}
{\sf E}[\Pi_{1,2}]&=&{\sf E}\left\{\left[F_{Y|W}({\sf Q}_\tau(W_{12})-B_{12}|W_{12})-F_{Y|W}({\sf Q}_\tau(W_{12}) |W_{12})\right]Z_{12}K_{h}(W_{12}-w)\right\}\notag\\
&\sim&-{\sf E}\left[f_{Y|W}({\sf Q}_\tau(W_{12}) |W_{12})Z_{12}B_{12}K_{h}(W_{12}-w)\right]\notag\\
&=&-\frac{1}{2} h^2f_{Y|W}({\sf Q}_\tau(w) |w)f_W(w)(1+o(1))\int s^{^\intercal} \frac{\partial^{2}{\sf Q}_{\tau}(w)}{\partial w \partial w^{^\intercal}}s \left(\begin{array}{c}
1\\
s
\end{array}\right) K(s)ds.\label{eqB.6}
\end{eqnarray}
The remainder in (\ref{eqB.6}) is $o(h^2)$. We next bound the variance of $\Pi_{1,2}$. Let ${\cal F}_X=\sigma\{X_1,\cdots,X_n\}$ and
\[
\xi_{ij}=[I(R_{ij}+B_{ij}\leq0)-I(R_{ij}\leq0)]Z_{ij}K_h(W_{ij}-w).
\]
Since $K(\cdot)$ has compact support and ${\sf Q}_\tau(\cdot)$ is twice continuously differentiable, $|B_{ij}|\leq Ch^2$ and $\|Z_{ij}\|\leq C$ whenever $K_h(W_{ij}-w)\neq0$. Then, by Assumption \ref{ass:2}(ii), we readily have that
\[
\big\|{\sf E}[\xi_{ij}\,|\,{\cal F}_X]\big\|\leq Ch^2|K_h(W_{ij}-w)|
\quad\text{and}\quad
{\sf E}\big[\|\xi_{ij}\|^2\,|\,{\cal F}_X\big]\leq Ch^2K_h^2(W_{ij}-w)\quad \textit{a.s.}
\]
As in (\ref{eqB.1})--(\ref{eqB.2}), the diagonal terms contribute $O(h^2(n^2h^{d_W})^{-1})$ to ${\sf Var}(\Pi_{1,2})$. Pairs with disjoint indices are independent with zero covariance. For pairs sharing exactly one index, say $(i,j)$ and $(i,k)$, using $\|\xi_{ik}\|\leq C|K_h(W_{ik}-w)|$, we obtain
\[
\big|{\sf Cov}(\xi_{ij},\xi_{ik})\big|\leq C\,{\sf E}\big[{\sf E}(\|\xi_{ij}\|\,|\,{\cal F}_X)\,|K_h(W_{ik}-w)|\big]+O(h^4)=O(h^{2-d_X}).
\]
Noting that ${\cal D}$ (defined in the proof of Lemma \ref{le:1} has cardinality of order $n^3$, we have, in general,
\begin{equation}\label{eqB.7}
{\sf Var}(\Pi_{1,2})=O\left(h^2(n^2h^{d_{W}})^{-1}+h^2(nh^{d_{X}})^{-1}\right)=o\left((nh^{d_{X}})^{-1}\right),
\end{equation}
which, together with (\ref{eqB.6}), completes the proof of (\ref{eqB.5}).

Under the condition (a) of Theorem \ref{Th1}(ii), $\sigma_n^2(w)\asymp(nh^{d_X})^{-1}$, which, together with (\ref{eqB.7}), gives 
\[
{\sf Var}(\Pi_{1,2})/(h^2+\sigma_n(w))^2=O(h^2)=o(1),
\] 
proving \eqref{eqB.5b}. Under the condition (b) of Theorem \ref{Th1}(ii), for pairs sharing exactly one index, say $(i,j)$ and $(i,k)$, $\xi_{ij}$ and $\xi_{ik}$ are conditionally independent given ${\cal F}_X$, so 
$$|{\sf Cov}(\xi_{ij},\xi_{ik})|\leq C\,{\sf E}\left[h^4|K_h(W_{ij}-w)K_h(W_{ik}-w)|\right]+O(h^4)=O\left(h^{4-d_X}\right).$$
Consequently, we obtain
$${\sf Var}(\Pi_{1,2})=O\left(h^2(n^2h^{d_W})^{-1}+h^4(nh^{d_X})^{-1}\right).$$ 
Since $(h^2+\sigma_n(w))^2\geq \min\{h^4, \sigma_n^2(w)\}$ and $\sigma_n^2(w)\asymp(n^2h^{d_W})^{-1}$ under (b), we may show that
$$
{\sf Var}(\Pi_{1,2})/(h^2+\sigma_n(w))^2=O\left(h^2+(nh^{d_X})^{-1}\right)=o(1),
$$
completing the proof of (\ref{eqB.5b}).

\end{proof}

\begin{lemma}\label{le:3}

Suppose that Assumptions \ref{ass:1}--\ref{ass:3} are satisfied. Let $\Pi_{2}$ be defined as in the proof of Theorem \ref{Th1}. Then we have 
\begin{equation}\label{eqB.8}
\Pi_{2}=\frac{1}{2}f_{Y|W}({\sf Q}_{\tau}(w)|w)f_{W}(w)\phi^{^\intercal}{\boldsymbol\Lambda}(K)\phi+o_P(1),
\end{equation}
where ${\boldsymbol\Lambda}(K)={\sf diag}\left\{1,\ \int ss^{^\intercal}K(s)ds\right\}$.

\end{lemma}

\begin{proof}[{\bf Proof of Lemma \ref{le:3}}] In order to complete the proof of (\ref{eqB.8}), it is sufficient to show 
\begin{equation}\label{eqB.9}
{\sf E}\left[\Pi_{2} | {\cal F}_X\right]=\frac{1}{2}f_{Y|W}({\sf Q}_{\tau}(w))f_{W}(w)\phi^{^\intercal}{\boldsymbol\Lambda}(K)\phi+o_P(1)
\end{equation}
and 
\begin{equation}\label{eqB.10}
\Pi_{2}-{\sf E}\left[\Pi_{2} | {\cal F}_X\right]=o_P(1),
\end{equation}
where ${\cal F}_X=\sigma\{X_1,\cdots,X_n\}$ denotes the $\sigma$-field generated by $X_1,\cdots X_n$.

We start with the proof of (\ref{eqB.9}). Note that 
\begin{eqnarray}
&&{\sf E}\left[I\left(R_{ij}+B_{ij} \leq u\right)- I\left(R_{ij}+B_{ij} \leq 0 \right)\ |\ {\cal F}_X\right]\notag\\
&=&F_{Y|W}({\sf Q}_\tau(W_{ij})-B_{ij}+u\ |\ W_{ij})-F_{Y|W}({\sf Q}_\tau(W_{ij})-B_{ij}\ |\ W_{ij})\notag\\
&=&u\left[f_{Y|W}({\sf Q}_{\tau}(w)\ |\ W_{ij})+o_P(1)\right],\notag
\end{eqnarray}
which indicates that
\begin{eqnarray}
{\sf E}\left[\Pi_{2} | {\cal F}_X\right]&=&\frac{1+o_P(1)}{n(n-1)}\sum_{1\leq i \neq j\leq n}K_{h}(W_{ij}-w)f_{Y|W}({\sf Q}_{\tau}(w)\ |\ W_{ij})\int_{0}^{Z_{ij}^{^\intercal}\phi} udu\notag\\
&=&\frac{1+o_P(1)}{2n(n-1)}\sum_{1\leq i \neq j\leq n}K_{h}(W_{ij}-w)f_{Y|W}({\sf Q}_{\tau}(w)\ |\ W_{ij})\phi^{^\intercal}Z_{ij}Z_{ij}^{^\intercal}\phi\notag\\
&=&\frac{1}{2}f_{Y|W}({\sf Q}_{\tau}(w)|w)f_{W}(w)\phi^{^\intercal}{\boldsymbol\Lambda}(K)\phi+o_P(1),\notag
\end{eqnarray}
completing the proof of (\ref{eqB.9}).

We next turn to the proof of (\ref{eqB.10}). Let 
\begin{eqnarray}
\Gamma_{ij}(\phi)&=&\int_{0}^{Z_{ij}^{^\intercal}\phi}\left[I\left(R_{ij}+B_{ij} \leq u\right)- I\left(R_{ij}+B_{ij} \leq 0 \right)\right]du-\notag\\
&&{\sf E}\left[\int_{0}^{Z_{ij}^{^\intercal}\phi}\left[I\left(R_{ij}+B_{ij} \leq u\right)- I\left(R_{ij}+B_{ij} \leq 0 \right)\right]du\ |\ {\cal F}_X\right]\notag\\
&=&\int_{0}^{Z_{ij}^{^\intercal}\phi}\left[I\left(R_{ij}+B_{ij} \leq u\right)- I\left(R_{ij}+B_{ij} \leq 0 \right)\right]du-\notag\\
&&{\sf E}\left[\int_{0}^{Z_{ij}^{^\intercal}\phi}\left[I\left(R_{ij}+B_{ij} \leq u\right)- I\left(R_{ij}+B_{ij} \leq 0 \right)\right]du\ |\ W_{ij}\right],\notag
\end{eqnarray}
and write
\[
\Pi_{2}-{\sf E}\left[\Pi_{2} | {\cal F}_X\right]=\frac{2}{n(n-1)}\sum_{i<j}\Gamma_{ij}^\circ(\phi,w)
\]
with
\[
\Gamma_{ij}^\circ(\phi,w)=\frac{1}{2}\left[\Gamma_{ij}(\phi)K_{h}(W_{ij}-w)+\Gamma_{ji}(\phi)K_{h}(W_{ji}-w)\right].
\]
Using the dyadic dependence property and following the proof of Lemma \ref{le:1}, letting ${\cal D}$ be defined after equation (\ref{eqB.1}), we may show that
\begin{eqnarray}
{\sf Var}\left\{\Pi_{2}-{\sf E}\left[\Pi_{2} | {\cal F}_X\right]\right\}&=&\frac{4}{n^2(n-1)^2}\sum_{i<j}{\sf Var}(\Gamma_{ij}^\circ(\phi,w))+\notag\\
&&\frac{4}{n^2(n-1)^2}\sum_{(i_1,j_1,i_2,j_2)\in{\cal D}}{\sf Cov}(\Gamma_{i_1j_1}^\circ(\phi,w), \Gamma_{i_2j_2}^\circ(\phi,w))\notag\\
&=&O\left(\left(n(n-1)h^{d_{W}}\right)^{-1}\right)+O\left(\left(nh^{d_{X}}\right)^{-1}\right)\notag\\
&=&O\left(\left(nh^{d_{X}}\right)^{-1}\right)=o(1),\label{eqB.11}
\end{eqnarray}
indicating that (\ref{eqB.10}) holds. 

\end{proof}

\begin{lemma}\label{le:4}

Suppose that Assumptions \ref{ass:1}--\ref{ass:3} are satisfied. Let $\Xi_{n,3}$ be defined as in the proof of Theorem \ref{Th1}(ii). Then we have 
\begin{equation}\label{eqB.12}
{\sf E}\left[\Xi_{n,3}^2\right]=o(1).
\end{equation}

\end{lemma}

\begin{proof}[{\bf Proof of Lemma \ref{le:4}}] Define
\begin{eqnarray}
M(X_i,X_j,U_i, U_j)&=&{\sf E}\left[\left(\theta^{^\intercal}\left\{T_{i,j}(w)-{\sf E}\left[T_{i,j}(w)|X_{i},X_{j},U_{i},U_{j}\right]\right\}\right)^2 |X_{i},X_{j},U_{i},U_{j}\right]-\notag\\
&&{\sf E}\left(\theta^{^\intercal}\left\{T_{i,j}(w)-{\sf E}\left[T_{i,j}(w)|X_{i},X_{j},U_{i},U_{j}\right]\right\}\right)^2.\notag
\end{eqnarray}
Note that
\begin{eqnarray}
\Xi_{n,3}&=&\frac{2}{n(n-1)\sigma_n^2(w|\theta)}\cdot\frac{2}{n(n-1)}\sum_{1\leq i<j\leq n}M(X_i,X_j,U_i, U_j)\notag\\
&=:&\frac{2}{n(n-1)\sigma_n^2(w|\theta)}\cdot \Xi_{n,3}^\dag,\label{eqB.13}
\end{eqnarray}
where $\Xi_{n,3}^\dag$ is a mean-zero U-statistic. Applying the Hoeffding-type decomposition as in (\ref{eqA.6}) and using the arguments in proving (\ref{eqA.7}), (\ref{eqA.8}) and (\ref{eqA.10}), we may show that
\begin{equation}\label{eqB.14}
{\sf E}\left[\left(\Xi_{n,3}^\dag\right)^2\right]=O\left(\frac{1}{nh^{d_X+2d_W}}+\frac{1}{n^2h^{3d_W}}\right),
\end{equation}
which, together with (\ref{eqA.11}) and (\ref{eqB.13}), leads to 
\begin{equation}\label{eqB.15}
{\sf E}\left[\Xi_{n,3}^2\right]=O\left(\left(nh^{d_X}\right)^{-3}+\left(nh^{d_X}\right)^{-4}\right)=o(1),
\end{equation}
as $nh^{d_X}\rightarrow\infty$ in Assumption \ref{ass:3}(ii). The proof is completed.

\end{proof}

\begin{lemma}\label{le:5}

(i) Suppose that Assumptions \ref{ass:1}, \ref{ass:2} and \ref{ass:3}(i)(ii) are satisfied, and that either (a) or (b) of Theorem \ref{Th1}(ii) holds. Let $\widehat\phi:=\widehat\phi(w)=\phi(\widehat{\sf Q}_\tau(w),\widehat{\sf Q}_\tau^\prime(w))$ and ${\bf V}=f_{Y|W}({\sf Q}_{\tau}(w)|w)f_{W}(w){\boldsymbol\Lambda}(K)$. Then
\begin{equation}\label{eqB.bahadur}
\widehat\phi=-{\bf V}^{-1}\Pi_1+o_P(\varrho_n)\quad \textit{with}\quad \varrho_n=h^2+\sigma_n(w).
\end{equation}

(ii) Under Assumptions \ref{ass:1}, \ref{ass:2} and \ref{ass:3}(i)(ii) only,
\begin{equation}\label{eqB.rate}
\widehat{\sf Q}_\tau(w)-{\sf Q}_\tau(w)=O_P\left(h^2+(nh^{d_X})^{-1/2}\right).
\end{equation}

\end{lemma}

\begin{proof}[{\bf Proof of Lemma \ref{le:5}}] Write $\phi=\varrho_ns$ and define
\[
G_n(s)=\varrho_n^{-2}\,\overline{L}_\tau(\varrho_ns)=U_n^{^\intercal}s+\varrho_n^{-2}\Pi_2(\varrho_ns),
\qquad U_n=\varrho_n^{-1}\Pi_1,
\]
where $\overline{L}_\tau(\cdot)$ and $\Pi_1$ are defined in the proof of Theorem \ref{Th1}(i), and $\Pi_2(\phi)$ is defined as $\Pi_2$ but with its dependence on $\phi$ made explicit. Note that $G_n(\cdot)$ is convex with $G_n(0)=0$ and is minimized at $\widehat s=\widehat\phi/\varrho_n$. By the basic corollary of \cite{HP11}, $\widehat s=-{\bf V}^{-1}U_n+o_P(1)$, which leads to (\ref{eqB.bahadur}), provided that (I) $U_n=O_P(1)$ and (II) for each fixed $s$, $\varrho_n^{-2}\Pi_2(\varrho_ns)=\frac12s^{^\intercal}{\bf V}s+o_P(1)$. Here ${\bf V}$ is positive definite by Assumptions \ref{ass:2} and \ref{ass:3}(i).

It follows from Lemma \ref{le:1} that $\Pi_{1,1}=O_P(\sigma_n(w))$ under the non-degeneracy case whereas ${\sf Var}(\Pi_{1,1})=O([n(n-1)h^{d_W}]^{-1})=O(\sigma_n^2(w))$ when the condition (b) of Theorem \ref{Th1}(ii) is satisfied under the degeneracy case.
Consequently, we obtain $\Pi_{1,1}=O_P(\sigma_n(w))$. On the other hand, by (\ref{eqB.5b}) in Lemma \ref{le:2}, $\Pi_{1,2}=O_P(h^2)+o_P(\varrho_n)$. Combining the above arguments, we readily have that
\[
U_n=\varrho_n^{-1}\Pi_1=\varrho_n^{-1}(\Pi_{1,1}+\Pi_{1,2})=O_P(1),
\]
verifying (I).

For (II), fix $s$ and let $t_{ij}=\varrho_nZ_{ij}^{^\intercal}s$. Then $|t_{ij}|\leq C\varrho_n$ whenever $K_h(W_{ij}-w)\neq0$. Decompose 
\[
\Pi_2(\varrho_ns)={\sf E}[\Pi_2(\varrho_ns)\,|\,{\cal F}_X]+\{\Pi_2(\varrho_ns)-{\sf E}[\Pi_2(\varrho_ns)\,|\,{\cal F}_X]\}.
\]
For the first term, we note that
\begin{eqnarray}
{\sf E}[\Pi_2(\varrho_ns)\,|\,{\cal F}_X]&=&\frac{1}{n(n-1)}\sum_{1\leq i\neq j\leq n}K_h(W_{ij}-w)\int_0^{t_{ij}}\big[F_{Y|W}({\sf Q}_\tau(W_{ij})-B_{ij}+u\,|\,W_{ij})-\nonumber\\
&&F_{Y|W}({\sf Q}_\tau(W_{ij})-B_{ij}\,|\,W_{ij})\big]du.\notag
\end{eqnarray}
and ${\sf Q}_\tau(W_{ij})-B_{ij}={\sf Q}_\tau(w)+(W_{ij}-w)^{^\intercal}{\sf Q}_\tau^\prime(w)$. By the joint continuity in Assumption \ref{ass:2}(ii), the integral equals $\frac12f_{Y|W}({\sf Q}_\tau(w)|w)t_{ij}^2(1+\epsilon_{ij})$, where $\max\{|\epsilon_{ij}|:K_h(W_{ij}-w)\neq0\}$ is bounded by a deterministic sequence tending to zero. Since $\frac{1}{n(n-1)}\sum_{i\neq j}|K_h(W_{ij}-w)|\,\|Z_{ij}\|^2=O_P(1)$, the contribution of the $\epsilon_{ij}$ terms is $o_P(1)$. Hence
\begin{eqnarray}
\frac{1}{\varrho_n^{2}}{\sf E}[\Pi_2(\varrho_ns)\,|\,{\cal F}_X]&=&\frac12f_{Y|W}({\sf Q}_\tau(w)|w)\,s^{^\intercal}\Big[\frac{1}{n(n-1)}\sum_{i\neq j}K_h(W_{ij}-w)Z_{ij}Z_{ij}^{^\intercal}\Big]s+o_P(1)\notag\\
&=&\frac12s^{^\intercal}{\bf V}s+o_P(1),\notag
\end{eqnarray}
since the bracketed average equals $f_W(w){\boldsymbol\Lambda}(K)+o_P(1)$ following the mean and variance calculations as in the proof of Lemma \ref{le:1}.

For the second term, let $\Gamma_{ij}=\Gamma_{ij}(\varrho_ns)$ be as in the proof of Lemma \ref{le:3}. Since $|\int_0^t[I(x\leq u)-I(x\leq0)]du|\leq|t|\,I(|x|\leq|t|)$ and $f_{Y|W}$ is bounded, we have 
\[
|\Gamma_{ij}|\leq|t_{ij}|I(|R_{ij}+B_{ij}|\leq|t_{ij}|)+Ct_{ij}^2,
\] 
so that 
\[
{\sf E}[\Gamma_{ij}^2\,|\,{\cal F}_X]\leq C\varrho_n^3\quad \text{and}\quad {\sf E}[|\Gamma_{ij}\Gamma_{ik}|\,|\,{\cal F}_X]\leq C\varrho_n^3
\]
on the support of the kernel weights. Proceeding as in (\ref{eqB.11}), the diagonal terms contribute $O(\varrho_n^3(n^2h^{d_W})^{-1})$ to the variance. For pairs with disjoint indices, $\Gamma_{ij}$ and $\Gamma_{kl}$ are conditionally independent given ${\cal F}_X$, as they depend on $(U_i,U_j,V_{ij})$ and $(U_k,U_l,V_{kl})$ respectively, and each has conditional mean zero, so their covariance vanishes. The pairs $(i,j)$ and $(j,i)$ contribute at the same order as the diagonal terms by the Cauchy-Schwarz inequality. Under (a), the pairs sharing exactly one index contribute $O(\varrho_n^3(nh^{d_X})^{-1})$. Under (b), the $\Gamma_{ij}$ are conditionally independent with conditional mean zero given ${\cal F}_X$, so these pairs do not contribute to the variance. Since $\sigma_n(w)\geq c(n^2h^{d_W})^{-1/2}$ always, and $\sigma_n(w)\geq c(nh^{d_X})^{-1/2}$ under (a),
\[
\frac{1}{\varrho_n^{4}}{\sf Var}\left(\Pi_2(\varrho_ns)-{\sf E}[\Pi_2(\varrho_ns)\,|\,{\cal F}_X]\right)=o(1),
\]
which completes the verification of (II) and thus the proof of (\ref{eqB.bahadur}).

We finally turn to the proof of (\ref{eqB.rate}). If neither (a) nor (b) is imposed, the same argument with $\varrho_n$ replaced by $\tilde\varrho_n=h^2+(nh^{d_X})^{-1/2}$ applies. By Lemma \ref{le:1}, $\Pi_{1,1}=O_P((nh^{d_X})^{-1/2})$; by (\ref{eqB.6}) and (\ref{eqB.7}), $\Pi_{1,2}=O_P(h^2)+o_P(\tilde\varrho_n)$; and the general bounds above give $\tilde\varrho_n^{-4}{\sf Var}(\Pi_2(\tilde\varrho_ns)-{\sf E}[\Pi_2(\tilde\varrho_ns)\,|\,{\cal F}_X])=O((nh^{d_X})^{-1/2})$. This proves (\ref{eqB.rate}).
\end{proof}

%%%%%%%%%%%%%%%%%%

\section*{Appendix C:\ Practical issues in the variance estimation}
\renewcommand{\theequation}{C.\arabic{equation}}
\setcounter{equation}{0}

The two components of $\sigma_n^2(w)$ reflect two sources of sampling variation: the dependence between dyads that share a node and the variation that remains at the dyad level. Theorem~\ref{Th1} retains both components and accommodates non-degeneracy and degeneracy without preliminary classification. Estimating these components in a finite sample is a separate issue. In particular, an estimated node component can be positive even when its population counterpart vanishes. This appendix explains the practical diagnostic used in Section~\ref{sec4.4} and the intuition behind the choice of variance component.

\smallskip

\noindent\underline{\bf Separating the node signal from estimation noise.}\ \ The covariance estimator in (\ref{eq3.12}) averages $[\widehat A_k^\circ(w)]^2$ across nodes. Each $\widehat A_k^\circ(w)$ is itself an average of $m=n-n_1-1$ weighted quantile scores. Its square therefore reflects both the underlying node signal and the sampling noise in estimating that signal. Even if the signal is zero, a noisy average will generally have a positive square. This contribution can obscure the behavior of the node signal as the bandwidth changes.

\smallskip

The adjustment subtracts an estimate of this noise contribution. With the bandwidths fixed, and conditional on the reference block and the observed and latent attributes of node $k$, the $m$ terms entering $\widehat A_k^\circ(w)$ are independent with a common distribution. The variance of their average is consequently the variance of an individual term divided by $m$. Let $\widehat s_k^2(w)$ be their sample variance, computed with divisor $m-1$. Subtracting $\widehat s_k^2(w)/m$ from the squared average motivates
\begin{equation}\label{eqC.1}
\widehat{\sf Cov}_{\rm adj}(w)=\max\left(0,\; \frac{1}{n-n_1}\sum_{k=n_1+1}^n \left\{\bigl[\widehat A_k^\circ(w)\bigr]^2-\frac{\widehat s_k^{\,2}(w)}{n-n_1-1}\right\}\right),
\end{equation}
where truncation is applied after averaging across nodes because the covariance being estimated is nonnegative. The purpose is to reduce the influence of estimation noise on the diagnostic, especially when the node signal is weak. Since the bandwidths are selected from the data, this argument provides motivation for the adjustment rather than an exact unbiasedness claim for the implemented estimator.

\smallskip

\begin{samepage}
We insert this adjusted covariance into (\ref{eq3.13}) to obtain the coefficient used for the diagnostic:
\begin{equation}\label{eqC.2}
\widehat\nu_{\tau,\mathrm{adj}}(w;h)
=4h^{d_X}\frac{\widehat{\sf Cov}_{\rm adj}(w;h)}
{\left[\widehat f_{Y|W}({\sf Q}_\tau(w))\widehat f_W(w)\right]^2}.
\end{equation}
\end{samepage}
The argument $h$ makes the bandwidth dependence explicit. The factor $h^{d_X}$ uses the same normalization as the main text. The adjustment is used only for the diagnostic: when the node component is selected for interval construction, we retain the original covariance estimator (\ref{eq3.12}) in $\widehat\nu_\tau(w)$.

\smallskip

\noindent\underline{\bf Reading the signal across bandwidths.}\ \ The diagnostic asks whether the adjusted coefficient is roughly stable as the bandwidth changes or instead grows with the width of the kernel window. Under non-degeneracy, the leading covariance signal has order $h^{-d_X}$, so the factor $h^{d_X}$ in (\ref{eqC.2}) leaves a coefficient that is approximately constant. Under degeneracy, there is no population node signal at the evaluation point. As a heuristic, an off-center contribution of order $h$ within the kernel window would yield a squared contribution of order $h^2$ after normalization. These two patterns suggest log--log slopes near zero and two, respectively. In our simulation setting, where $d_X=1$, a cutoff of one separates the two benchmark patterns.

\smallskip

We evaluate (\ref{eqC.2}) at $h=\gamma\widehat h_{\rm cv}$ for $\gamma\in\{0.75,1.00,1.25,1.50,1.75,2.00\}$. The reference quantile estimated on ${\mathscr M}_1$ at $\widehat h_{\rm US}$ and the density estimates are held fixed, so the comparison follows changes in the kernel bandwidth. We regress $\log\widehat\nu_{\tau,\mathrm{adj}}(w;h)$ on $\log h$, evaluating the normalization $h^{d_X}$ at each candidate bandwidth. Only positive finite coefficients enter the regression, since zero coefficients cannot be logged. A finite fitted slope below one selects the node component; a slope at or above one selects the dyad component. We require at least three usable points and use the dyad component when this requirement is not met or the fitted slope is nonfinite.

\smallskip

The resulting rule retains the component suggested to dominate in the sample. Writing $I_\tau(w)=1$ when a usable slope satisfies $\widehat s_\tau(w)<1$, and $I_\tau(w)=0$ otherwise, the variance estimate is
\begin{equation}\label{eqC.3}
\widehat\sigma_n^2(w)=
I_\tau(w)\frac{\widehat\nu_\tau(w)}{n\widehat h_{\rm US}^{d_X}}
+\{1-I_\tau(w)\}\frac{\widehat\nu_\tau^\dag(w)}{n(n-1)\widehat h_{\rm US}^{d_W}},
\end{equation}
where both coefficients are evaluated at the inference bandwidth and the selected density bandwidths. Thus the node component uses the original covariance, and the dyad component uses the density-based estimator in Section~\ref{sec3.2.1}. This choice follows the leading-term intuition: the node component dominates under non-degeneracy, whereas the dyad component governs the variance when the population node contribution vanishes.

\smallskip

The slope comparison is a practical diagnostic, not a formal test of degeneracy. Its benchmark patterns provide intuition rather than a consistency result, particularly because the reference quantile is estimated. The coverage results in Section~\ref{sec4.4} assess this finite-sample implementation. The population adaptation result does not by itself establish relative consistency of the estimated variance, and a general theoretical justification of the switching rule remains an open question for future research.

%%%%%%%%%%%%%%%%%%%%%%%%%%%%%%%

%%%%%%%%%%%%%%%%%%%%%%%%%%%%%%%%%%%%%%%%%%%%%%%%%%%%%%%%%%%%%%%%%%%

\end{document}